\documentclass[fleqn,usenatbib]{mnras}

\usepackage{newtxtext,newtxmath}
\usepackage{xcolor}
\usepackage{subfig}
\usepackage{lineno}
\usepackage[T1]{fontenc}
\DeclareRobustCommand{\VAN}[3]{#2}
\let\VANthebibliography\thebibliography
\def\thebibliography{\DeclareRobustCommand{\VAN}[3]{##3}\VANthebibliography}

\newcommand{\cc}[1]{\textcolor{orange}{#1}}

\newcommand{\g}[1]{\textcolor{gray}{#1}}

\usepackage[normalem]{ulem}

\usepackage{graphicx}	
\usepackage{amsmath}	
\usepackage{xspace}
\usepackage{comment}
\usepackage{eso-pic}%
\newcommand{\redmagic}{\textsc{redMaGiC}\xspace}
\newcommand{\maglim}{\textsc{MagLim}\xspace}
\newcommand{\Planck}{\textit{Planck}\xspace}

\title[DES Y3 $\times$ $y$-SPT]{Mapping gas around massive galaxies: cross-correlation of DES Y3 galaxies and Compton-$y$-maps from SPT and {\it Planck}}

\author[J.~S\'{a}nchez et al.]{
\parbox{\textwidth}{
\Large
J.~S\'{a}nchez,$^{1,2,3}$
Y.~Omori,$^{4,3,5,6}$
C.~Chang,$^{4,3}$
L.~E.~Bleem,$^{7,3}$
T.~Crawford,$^{4,3}$
A.~Drlica-Wagner,$^{4,2,3}$
S.~Raghunathan,$^{8}$
G.~Zacharegkas,$^{4,3}$
T.~M.~C.~Abbott,$^{9}$
M.~Aguena,$^{10}$
% ...
\begin{center} (DES and SPT Collaborations) \end{center}
\emph{\normalsize Affiliations are listed at the end of the paper}
}
}

\date{Accepted XXX. Received YYY; in original form ZZZ}

\pubyear{2022}

\begin{document}
\AddToShipoutPictureBG*{%
  \AtPageUpperLeft{%
    \hspace{0.75\paperwidth}%
    \raisebox{-4.5\baselineskip}{%
      \makebox[0pt][l]{\textnormal{FERMILAB-PUB-22-603-PPD}}
}}}%

\label{firstpage}
\pagerange{\pageref{firstpage}--\pageref{lastpage}}
\maketitle

\begin{abstract}
We cross-correlate positions of galaxies measured in data from the first three years of the Dark Energy Survey with Compton-$y$-maps generated using data from the South Pole Telescope (SPT) and the {\it Planck} mission. We model this cross-correlation measurement together with the galaxy auto-correlation to constrain the distribution of gas in the Universe. We measure the hydrostatic mass bias or, equivalently, the mean halo bias-weighted electron pressure $\langle b_{h}P_{e}\rangle$, using large-scale information. We find $\langle b_{h}P_{e}\rangle$ to be $[0.16^{+0.03}_{-0.04},0.28^{+0.04}_{-0.05},0.45^{+0.06}_{-0.10},0.54^{+0.08}_{-0.07},0.61^{+0.08}_{-0.06},0.63^{+0.07}_{-0.08}]$ meV cm$^{-3}$ at redshifts $z \sim [0.30, 0.46, 0.62,0.77, 0.89, 0.97]$. These values are consistent with previous work where measurements exist in the redshift range. We also constrain the mean gas profile using small-scale information, enabled by the high-resolution of the SPT data. We compare our measurements to different parametrized profiles based on the cosmo-OWLS hydrodynamical simulations. We find that our data are consistent with the simulation that assumes an AGN heating temperature of $10^{8.5}$K but are incompatible with the model that assumes an AGN heating temperature of $10^{8.0}$K. These comparisons indicate that the data prefer a higher value of electron pressure than the simulations within $r_{500c}$ of the galaxies' halos.

\end{abstract}


\section{Introduction}
\label{sec:intro}

In the canonical cold dark matter paradigm, dark matter  collapses non-linearly to form gravitationally bound halos. These dark matter halos host the visible baryonic matter, dominantly in the form of stars and hot gas. The interactions of these three components -- dark matter, stars, and gas -- are critical for understanding the astrophysical processes that govern galaxy formation and evolution \citep[see][for a review]{Naab2017}. Constraining these astrophysical processes is also one of the main focuses in galaxy surveys today \citep{Semboloni2013,Eifler2015, Huang2019, Chisari2019,Chen2022}, as astrophysical uncertainties are becoming the limiting factor for cosmological analyses.

The joint analysis of observations at optical/near-infrared and microwave wavelengths provides an emerging opportunity to study the connection between gas, stars (which form the visible part of the galaxies), and dark matter. 
Surveys at optical/near-infrared wavelengths precisely measure the locations and properties of hundreds of millions of galaxies. 
On the other hand, at microwave wavelengths, the thermal Sunyaev-Zeldovich effect \citep[tSZ,][]{1972CoASP...4..173S}, which arises from the inverse Compton scattering of cosmic microwave background (CMB) photons with high-energy electrons, effectively traces the distribution of hot gas in the Universe. The amplitude of this effect depends on the average thermal pressure profile of galaxy groups and clusters~\citep{Battaglia2012, Bhattacharya2012}. The observable quantity of the tSZ effect is usually measured via the Compton-$y$ parameter, which we will refer to as $y$ throughout this paper.  
As shown in \citet{Hill2018} and \citet{Pandey2019}, on large scales, the cross-correlations between galaxies and maps of $y$ can constrain the thermal energy content of the Universe.
In addition, \citet{Schaan2021} showed that the small-scale information of the galaxy-$y$ correlation further constrains the gas thermodynamics and the properties of astrophysical feedback processes. On the practical side, these cross-correlation measurements are very robust to systematic effects, as the two datasets are almost completely independent and experimental systematic effects that only reside in one dataset will not bias our final measurement. The robustness of these cross-correlation measurements makes them very attractive as we approach an era where observational cosmology is increasingly systematics-limited.


Most recent analyses have focused on constraining the thermodynamics of intergalactic gas through measurements of the quantity $\langle b_{h}P_{e}\rangle$, often referred to as the halo bias-weighted mean electron pressure.\footnote{$b_{h}$ refers to the halo bias, which is used in the calculation of $\langle b_{h}P_{e}\rangle$ as shown in Equation~\ref{eq:bPe}.} This quantity is interesting for two reasons: i) it directly probes the mean electron pressure of the Universe and the growth of structure, ii) it is the direct observable of the SZ-galaxy cross-correlation at large scales~\citep{Vikram2017}. Some of these analyses then rely on the halo model to obtain the so-called hydrostatic bias (or mass bias), usually denoted $b_{H}$~\citep{Vikram2017, Pandey2019, Koukoufilippas2020, Yan2021}.

Three previous studies have combined the galaxy-$y$ cross-correlation with the the galaxy-galaxy auto-correlation \citep{Vikram2017,Pandey2019,Koukoufilippas2020}. All three of these studies focused mainly on the \Planck MILCA $y$-map~\citep{Planck2016} but used different galaxy samples: \citet{Vikram2017} used the group catalog from SDSS produced by \citet{Yang2007} (halo mass $10^{11}$--$10^{15}$M$_{\odot}/h$, redshift range 0.01--0.2); \citet{Pandey2019} used DES \redmagic galaxies from \citet{Rozo2016} (halo mass $\sim$ 10$^{13}$M$_{\odot}/h$, redshift range 0.15--0.75), and \citet{Koukoufilippas2020} used a combination of 2MASS \citep{Bilicki2014} and WISE $\times$ SuperCOSMOS \citep{Bilicki2016} galaxies (halo mass $\sim 10^{12}$--$10^{13}$M$_{\odot}/h$, redshift range $<$0.4). The first two measurements were carried out in real-space while the last one used a harmonic-space estimator. The combination of the galaxy-$y$ cross-correlation and the galaxy auto-correlation breaks the degeneracy between $\langle b_{h}P_{e}\rangle$ and the galaxy bias. All three analyses show consistent results, albeit with large uncertainties. 

Focusing mainly on the MILCA $y$-map, \citet{Yan2021} took a different approach and combined the galaxy-$y$ cross-correlation and cross-correlation of galaxy and CMB lensing, arguing that a pure cross-correlation analysis is less prone to systematic effects. The galaxy sample used is the weak lensing sample in the KiDS-1000 dataset (halo mass $\sim 10^{12} - 10^{15} M_{\odot}/h$, redshift $<1$). 

\citet{Chiang2020} explored a rather different route by using the individual temperature maps from different frequencies in \Planck instead of using a $y$-map. They cross-correlate these maps with spectroscopic galaxies in SDSS spanning a large range in mass and redshift (halo mass $\sim 10^{11.5} - 10^{13.5} M_{\odot}/h$, redshift $<3$). The cross-correlation amplitudes of the different maps at a given redshift effectively gives a Spectral Energy Distribution (SED), which the authors fit to a combination of the tSZ and cosmic infrared background (CIB) SEDs. The result from this multi-channel SED fitting approach can directly be translated into $\langle b_{h}P_{e}\rangle$. \citet{Chiang2020} showed that this procedure yields less CIB contamination at high redshift than using the $y$-maps directly. 


All the aforementioned analyses rely mainly on large-scale information, as they are limited by the resolution of the \Planck maps at $\sim10$ arcmin. With the help of high-resolution $y$-maps from ongoing CMB experiments, we are starting to explore the gas distributions on smaller scales. \citet{Schaan2021} cross-correlated the \textsc{CMASS} (mean redshift $\langle z\rangle=0.55$, mean virial mass $\langle M_{\rm vir}\rangle=2.1\times 10^{13}$M$_{\odot}/h$) and \textsc{LOWZ} ($\langle z\rangle=0.31$, $\langle M_{\rm vir}\rangle=3.5\times 10^{13}$M$_{\odot}/h$) samples from Baryon Oscillation Spectroscopic Survey \citep[BOSS,][]{Dawson2013} DR10 and DR12 with the $y$-map constructed by combining Atacama Cosmology Telescope \citep[ACT,][]{Swetz2011} DR5 and {\it Planck} data~\citep{Planck2016}. The tSZ profiles were measured at high significance within the size of a typical halo. \citet{Amodeo2021} compared these measured profiles to those in hydrodynamical simulations and found the measured profiles to disagree with simulations especially at large radii -- the data shows a peakier profile in general. This could suggest that the sub-grid stellar and AGN feedback models in these simulations 
do not sufficiently heat the gas in the outer regions. \citet{Gatti2021} and \citet{Pandey2021} recently performed a cross-correlation of galaxy weak lensing with the ACT DR4 tSZ maps. 
They showed that, when using the gas model in \citet{LeBrun2015}, the shear-$y$ measurements slightly prefer a stronger AGN feedback model than cosmic-shear-only analyses. \citet{Troester21} also found that shear-$y$ prefers a stronger AGN feedback than cosmic shear~\citep{Troester21KIDS}. Other constraints on small-scale baryonic effects on cosmic shear measurements include \citet{Yoon2021}, \citet{Huang2021}, \citet{Chen2022b}, \citet{Chen2022}, and \citet{Schneider2022}.

Building on past work (e.g., \citealt{Vikram2017,Pandey2019,Koukoufilippas2020}), we measure and model two sets of correlation functions: galaxy clustering, ``$gg$'', and the galaxy-$y$, ``$gy$'' cross-correlation. We use a galaxy sample defined for the recent large-scale structure cosmology analysis from the Dark Energy Survey \citep[DES,][]{DES2021} the \maglim galaxies,  the MILCA $y$-maps from \citep{Planck2016}, as well as $y$-maps constructed using a combination of South Pole Telescope (SPT) and \Planck data from \citet{Bleem2021}. In particular, the cross-correlations between DES galaxies and SPT-SZ+{\it Planck} $y$-maps have several advantages relative to previous analyses: (1) the galaxies are well-characterized and vetted for observational systematics (2) the galaxies span a large redshift range (up to $z \sim 1$) with good statistics reaching small scales (3) the $y$-map has lower noise and higher resolution compared to $y$-maps produced using {\it Planck} data only. Finally, as this galaxy sample is used for cosmological analyses in DES, our results can help constrain astrophysical nuisance parameters and improve cosmological constraints from DES. We first look at the two-halo regime (large-scales) to constrain $\langle b_{h}P_{e}\rangle$ and compare with previous work. We then look into the one-halo regime to map the gas profile around galaxies in our sample. We note also that our galaxy sample has been extensively characterized via galaxy-galaxy lensing in an HOD analysis \citep{Zacharegkas2021} and has been determined to have an average halo mass of a few times $10^{13}$M$_{\odot}/h$.

This paper is structured as follows. In Section~\ref{sec:methods} we describe the theoretical prescription used to model the distributions of galaxies and gas.
In Section~\ref{sec:data_sample} we describe the galaxy and $y$-maps that we use in this analysis.
In Section~\ref{sec:analysis} we describe our analysis prescription including our covariance matrix, scale cuts, likelihood inference, and systematic tests.
In Section~\ref{sec:results1} we discuss our constraints on the hydrostatic mass bias using large-scale information, and in Section~\ref{sec:results2} we discuss our constraints on gas profiles using small-scale information.
We conclude in Section~\ref{sec:conclusions}.


\section{Theory}
\label{sec:methods}

The goal of this analysis is to use the HOD framework to model the angular galaxy-galaxy auto spectrum $C^{\rm gg}_{\ell}$ and the galaxy-$y$ cross spectrum $C^{\rm gy}_{\ell}$. This requires a number of components that we detail in the following subsections: (1) the 3D galaxy auto spectrum and galaxy-$y$ cross spectrum following the Halo Model framework (Section~\ref{ssec:power_spectrum}) (2) an HOD model that describes how galaxies are associated with halos (Section~\ref{ssec:HOD}) (3) a functional form for the electron pressure profile around galaxies (Section~\ref{ssec:profile}), and (4) projecting the 3D galaxy-galaxy and galaxy-$y$ cross spectrum to 2D where our measurements are made, while incorporating a number of observational systematic effects (Section~\ref{ssec:projection}).  

Throughout, we will use $\delta_{g}(\hat{n})$ to represent the projected galaxy density and $y(\hat{n})$ to represent the $y$ signal at a given direction $\hat{n}$. The projected galaxy density can be expressed as,
\begin{equation}
    \delta_{g}(\hat{n}) = \int dz n_{g}(z) \Delta_{g}(\chi(z), \hat{n}),
\label{eq:g_hat}
\end{equation}
where $n_{g}(z)$ is the normalized redshift distribution of the galaxies that integrates to unity, $\chi(z)$ the comoving distance at redshift $z$, and $\Delta_{g}$ is the 3D overdensity of number of galaxies. 
The $y$ signal is expressed as,
\begin{equation}
y(\hat{n}) = \frac{\sigma_{T}}{m_{e}c^{2}}\int \frac{d\chi}{1+z} P_{e}(\chi(z), \hat{n}),   
\label{eq:y_hat}
\end{equation}
where $\sigma_{T}$ is the Thomson scattering cross-section, $m_{e}$ the electron mass, and $P_{e} = n_{e} T_{e}$ is the electron pressure. 

Unless explicitly stated, during this study we use the best-fit cosmological parameters from \textit{Planck}~\citep{Planck2018Cosmo}: $\Omega_{c}=0.2607,\ \Omega_{b}=0.0489,\ h=0.6766,\ n_{s}=0.9665$, and $\sigma_{8}=0.8102$. We choose to adopt this cosmology to make comparisons with previous analyses; we have verified that changing to the DES Y3 cosmology~\citep{DES2021} does not alter the main conclusions of this paper. 



\subsection{Halo model}
\label{ssec:power_spectrum}
We describe the 3D galaxy auto spectrum and galaxy-$y$ cross spectrum using the halo model~\citep[see][and references therein]{2000MNRAS.318..203S,2000MNRAS.318.1144P}. Both power spectra include two contributions: the 1-halo term, $P^{1h}(k)$, which describes the distribution of galaxies or $y$ inside a halo; and the 2-halo term, $P^{2h}(k)$, which describes the spatial distribution of the halos themselves. The profiles are described in Fourier space, and therefore are functions of the Fourier modes $k$ instead of radius $r$. For each two sets of observables $u$ and $v$ (which could be dark matter $m$, galaxy density $\delta_{g}$, or tSZ $y$) we assume that their radial distribution inside the halo follow the profiles $U(r | M)$ and $V(r | M)$, where $M$ is the halo mass and $r$ is the distance from the halo center. Their profiles in Fourier space, 
assuming isotropy, can be written as,  
\begin{equation}
U(k | M) = 4\pi \int_{0}^{\infty} r^{2}dr \frac{\sin{(kr)}}{kr} U(r | M).
\end{equation}
We can then write the cross power spectrum of the two observables, 
$P_{uv}(k)$, using the halo model as follows:
\begin{equation}
P_{uv}(k) = P^{1h}_{uv}(k) + P_{uv}^{2h}(k),
\label{eq:PS_total}
\end{equation}
with
\begin{equation}
P^{1h}_{uv}(k) = \int dM \frac{dn}{dM} \langle U(k | M) V (k | M) \rangle,    
\end{equation}
where $\langle \dots \rangle$ denote the ensemble average, and 

\begin{equation}
P^{2h}_{uv}(k) = \langle b_{h} U \rangle \langle b_{h} V \rangle P_{L}(k),     
\end{equation}
where
\begin{equation}
\langle b_{h} U \rangle = \int dM \frac{dn}{dM} b_{h}(M)\langle U(k |M)\rangle.
\label{eq:bu_u}
\end{equation}
Here, $b_{h}$ is the halo bias, $\frac{dn}{dM}$ is the halo mass function, and $P_{L}(k)$ is the linear matter power spectrum. We use the halo mass function from~\citet{2010ApJ...724..878T}, and the linear power spectrum from \texttt{CAMB}~\citep{Lewis:1999bs}. 

In order to correct for inaccuracies in the 1-to-2-halo transition~\citep{Mead15}, we follow~\citet{Koukoufilippas2020, Nicola2020} and modify the halo model power spectrum by multiplying it by the scale-dependent ratio,
\begin{equation}
R(k) = \frac{P_{\rm{HaloFit}}(k)}{P_{\rm{Halo Model}}(k)},    
\end{equation}
where $P_{\rm{HaloFit}}(k)$ is the HaloFit power spectrum from~\citet{2012ApJ...761..152T} and $P_{\rm{Halo Model}}(k)$ is the halo model matter power spectrum described in this section, i.e., Equation~\ref{eq:PS_total} assuming that the observable is dark matter. Thus, the final 3D power spectra for both the galaxy auto-spectrum and the galaxy-$y$ cross spectrum take the form
\begin{equation}
    P^{\rm final}_{uv}(k) = R(k) P_{uv}(k),
\end{equation}
where $u=v=\delta_{g}$ for the galaxy auto spectrum and $u=\delta_{g}$, $v=y$ for galaxy-$y$ cross spectrum. We will drop the super script ``final'' for the remainder of this text for simplicity.  

\subsection{Halo Occupation Distribution modeling}
\label{ssec:HOD}
In order to model galaxy density profiles, we use an HOD model \citep[][]{2000MNRAS.318.1144P, 2002PhR...372....1C, 2002ApJ...575..587B, 2005ApJ...633..791Z}. In this model, dark matter halos are populated by central and satellite galaxies. We assume that central galaxies follow a Bernouilli distribution while satellites follow a Poisson distribution \citep{2013MNRAS.430..725V}. Typically, the halo radii are defined as the size of a sphere containing a mass $M_{\Delta}$,
\begin{equation}
    M_{\Delta} = \frac{4\pi}{3}\rho_{*}(z)\Delta r^{3}_{\Delta}.
\end{equation}
We follow previous works~\citep[e.g.,][]{2010A&A...517A..92A, 2016A&A...594A..24P, 2018MNRAS.477.4957B, Koukoufilippas2020} and choose $\rho_{*} = \rho_{c}$ and $\Delta=500$ following the rederived concentration-mass relation from~\citet{2008MNRAS.390L..64D}:
\begin{equation}
    c_{500c}(M, z) = A(M/M_{\rm pivot})^{B}(1+z)^{C},
\end{equation}
with $M_{\rm pivot}=2.7 \times 10^{12} M_{\odot}$ and $(A, B, C) = (3.67, -0.0903, -0.51)$.

In our HOD model, the average number of central galaxies for a halo of mass $M$ is modeled as:
\begin{equation}
    \langle N_{c}(M) \rangle = \frac{1}{2} \left[ 1 + \mathrm{erf} \left(\frac{\log{\left(M/M_{\rm min}\right)}}{\sigma_{\rm{ln}M}}\right)\right],
    \label{eq:n_centrals}
\end{equation}
where $M_{\rm min}$ corresponds to the characteristic minimum mass of halos, and $\sigma_{\rm{ln}M}$ is the halo mass dispersion. Assuming that the halo can only form satellites if its mass is larger than a certain threshold, $M_{0}$, the average number of satellites can be described as:
\begin{equation}
    \langle N_{s}(M) \rangle = N_{c}(M) \Theta (M - M_{0}) \left(\frac{M-M_{0}}{M'_{1}}\right)^{\alpha_{s}}.
\end{equation}
Following~\citet{Ando2018, Koukoufilippas2020} we choose $M_{\rm min}=M_{0}$. As a consequence, all halos containing one or more satellites contain one central, located at the center of its parent halo.

For the satellites, we use the parametrization from~\citet{Ando2018} where the satellite galaxies follow a NFW profile~\citep{1996ApJ...462..563N} with a characteristic scale radius $\beta_{g}$, up to a maximum radius $\beta_{\rm max}$:
\begin{equation}\label{eq:NFW_prof}
\begin{split}
    U_{s}(r | M)  & \propto \Theta(\beta_{\rm max} - r)\left(\frac{1}{r/\beta_{g}(1+r/\beta_{g})^{2}}\right).
\end{split}
\end{equation}

\subsection{tSZ gas profile}
\label{ssec:profile}

For the electron pressure profile, we
use a generalized NFW (GNFW) profile~\citep{Nagai2007, 2010A&A...517A..92A}. In real space, it takes the form:
\begin{equation}
P_{e}(r) = P_{*}p(r/r_{500c}),
\label{eq:gas_prof}
\end{equation}
where 
the GNFW form factor, $p(r/r_{500c})$, is:
\begin{equation}
p(x) = (c_{P}x)^{-\gamma}\left[1+(c_{P}x)^{\alpha}\right]^{\frac{\gamma-\beta}{\alpha}}.     
\end{equation}
We choose the initial values $(\alpha, \beta, \gamma, c_{P}) = (1.33, 4.13, 0.31, 1.81)$ as found by~\citet{Planck2013a}. In parts of our analysis, we allow these parameters to float in our fits, which allows us to extract additional information about the gas profile. The normalization parameter $P_{*}$ is: 
\begin{equation}
P_{*} =P_{0} \left(1.65 \, \rm{eV \, cm}^{-3}\right) h_{70}^{8/3} \left(\frac{h_{70}(1-b_{H})M_{500c}}{3\times 10^{14}M_{\odot}}\right)^{0.79},
\label{eq:P_star}
\end{equation}
with $P_{0}=6.41$ the normalization constant, $h_{70} = H_{0}/(70 \, \rm{km \, s^{-1} \, Mpc^{-1}})$, 
and $b_{H}$ is the so-called hydrostatic mass bias, which corresponds to the fractional bias in the inferred mass using a gas proxy, assuming hydrostatic equilibrium with respect to its true mass. 
In the literature, this quantity is often expressed in terms of $B=\frac{1}{1-b_{H}}$. This parameter is typically used to calibrate the relationship between the mass inferred by the hot gas pressure profiles and the halo mass. Accurately calibrating this quantity can help to improve the calibration for cosmological studies using clusters~\citep{McClintock19, Miyatake19}.

The bias-weighted thermal energy (or bias-weighted electron pressure of the Universe) is then given by~\citep{Vikram2017, Pandey2019},
\begin{equation}
\langle b_h P_{e} \rangle = \int dM \frac{dn}{dM} b_{h}(M) \int_{0}^{\infty} dr 4\pi r^{2} P_{e}(r, M),
\label{eq:bPe}
\end{equation}
where $b_{h}$ is the halo bias, as in Equation~\ref{eq:bu_u}. We note that unlike Equation~\ref{eq:bu_u}, $\langle b_h P_{e} \rangle$ here is not a function of scale.
$\langle b_h P_{e} \rangle$ is closely related to the hydrostatic mass bias $b_H$, which is often what is shown in literature (or more precisely, $1-b_{H}$). 

\subsection{Projecting to the observable space}
\label{ssec:projection}
In this work we use the projected galaxy density $\delta_{g}(\hat{n})$, and the projected $y$-signal, $y(\hat{n})$. In general, a projected field $u(\hat{n})$ is related to its 3D field, $U$ via
\begin{equation}
    u(\hat{n}) = \int d\chi W_{u}(\chi(\hat{n}))U(\chi(\hat{n})),
\end{equation}

\noindent where $W_{u}(\chi(\hat{n}))$ is the window function, or radial kernel of the field. In the case of $\delta_{g}, y$, from Equations~\ref{eq:g_hat},\ref{eq:y_hat} their radial kernels take the form:
\begin{equation}
    W_{g}(\chi)=\frac{H(\chi)}{c}n_{g}(z(\chi));
    \label{eqn:win_gal}
\end{equation}
\begin{equation}
    W_{y}(\chi)=\frac{\sigma_{T}}{m_{e}c^{2}}\frac{1}{1+z(\chi)}.
    \label{eqn:win_y}
\end{equation}
In general, given two fields, $A, B$, their angular power spectrum $C_{\ell}^{AB}$ can be related to their 3D power spectrum $P_{AB}$:
\begin{equation}
    C^{AB}_{\ell} = \int \frac{d\chi W_{A}(\chi)W_{B}(\chi)}{\chi^{2}}P_{AB}\left(k=\frac{\ell+1/2}{\chi}, z(\chi)\right),
    \label{eqn:c_ell}
\end{equation}
where we use the Limber approximation~\citep{Limber1953}, which is sufficient given our smooth radial kernels, and the range of scales $(\ell > 150)$ that we consider in our work \citep{Fang2020}. In our study, we focus on the measurement and modeling of $C_{\ell}^{gg}$, and $C_{\ell}^{gy}$, using the relevant 3D power spectra defined in previous subsections. The power spectra models are computed using the Core Cosmology Library package, \texttt{CCL}~\citep{Chisari2019}, and are subsequently smoothed in order to account for pixelation and other smoothing in the input maps. In particular, we multiply $C_{\ell}^{gg}$ by the corresponding \textsc{HEALPix} window function, and $C_{\ell}^{gy}$ by the combination of the \textsc{HEALPix} window function of the maps involved and a Gaussian kernel with the FWHM of the $y$-map under consideration. 

We consider a shift parameter, $\Delta z_{i}$, on the original photometric redshift distribution of the galaxies for the bin $i$, $n_{pz, i}(z)$, and consider an additional stretch parameter, $\sigma_{z,i}$, as nuisance parameters following other DES analyses~\citep{DES2021, 2020arXiv201212826C}.
Thus, the resulting redshift distribution of galaxies for bin $i$, $n_{g,i}(z)$, is given by
\begin{equation}
    n_{g,i}(z) = \frac{1}{\sigma_{z,i}}n_{pz, i}\left(\frac{z-z_{{\rm mean}, i}-\Delta z_{i}}{\sigma_{z,i}}+z_{{\rm mean},i}\right),
\end{equation}
where $z_{{\rm mean}, i}$ is the mean redshift of the original photo-$z$ distribution $n_{pz, i}(z)$. We use the same priors for the photo-$z$ nuisance parameters as the aforementioned analyses. Plugging $n_{g,i}$ in Equation \ref{eqn:win_gal} we can obtain the galaxies' window function and, in combination with Equations~\ref{eqn:win_y} and~\ref{eqn:c_ell}, a model for $C^{gg}_{\ell}$ and $C^{gy}_{\ell}$.

\section{Data}
\label{sec:data_sample} 

Below we provide a brief description of the data products used in this work -- in particular the galaxy sample and the $y$-maps. The galaxy data products have been separately tested extensively in other studies~\citep{DES2021,Rodriguez-Monroy2021,Porredon2021,Pandey2021,Zacharegkas2021}, and  the $y$-maps has been thoroughly tested in \citet{Bleem2021} and \citet{Planck2016}. 

\subsection{DES Y3 galaxy sample}
We use data from DES~\citep[][]{DES:2005}, which collected data using the Dark Energy Camera~\citep[DECam,][]{2015AJ....150..150F} during 6 years observation at the Cerro Tololo Inter-American Observatory (CTIO). DES surveyed $\sim 5000$ square degrees of the southern sky using 5 broadband filters ($grizY$). In this work, we utilize data from the first 3 years of DES observing (Y3; 2013-2016). In particular, we use the \maglim galaxies~\citep{y3-2x2maglimforecast}. The main difference in this work compared to other Y3 studies is that we use the small-scale galaxy clustering measurements that are not used in most of the cosmological studies \citep[][looked at small scales, but only in galaxy-galaxy lensing]{Zacharegkas2021}. Since the SPT-SZ survey only overlaps the southern part of the DES footprint (${\rm Dec} < -40^{\circ}$), the galaxy sample is further split into a `northern' and `southern' regions, that we separately cross-correlate with different $y$-maps.

\begin{figure}
    \centering
    \includegraphics[width=0.99\columnwidth]{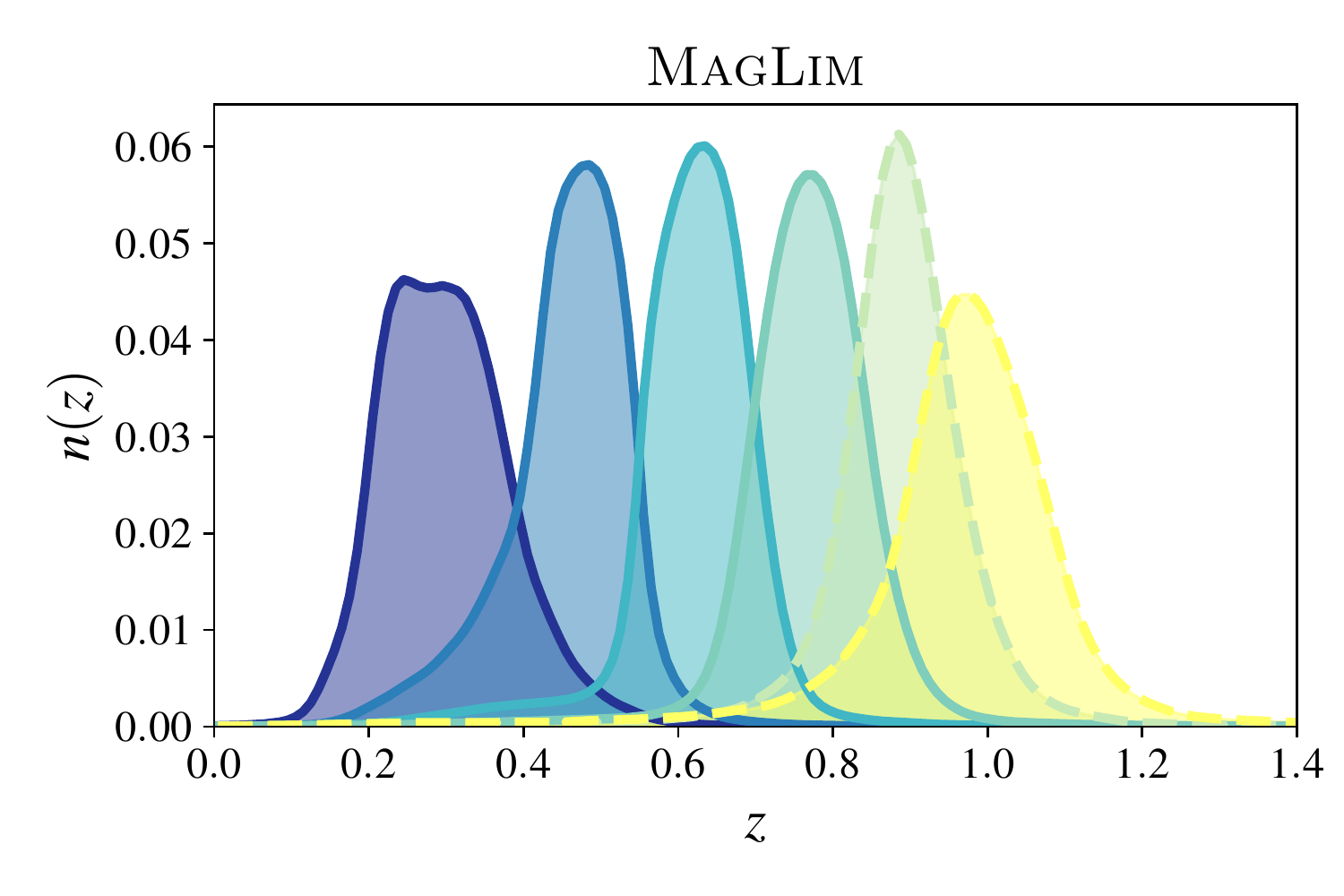}
    \caption{Redshift distributions of the fiducial (solid lines) and non-fiducial (broken lines) galaxy samples used in this study. The distributions are normalized so that their integral is 1. Details about the calibration of these distributions can be found in~\citet{2020arXiv201212826C}.}
    \label{fig:zdists}
\end{figure}

The \maglim sample is defined with a $i$-band magnitude cut that evolves linearly with the photometric redshift estimate: $i < a z_{\rm phot} + b$, where $z_{\rm phot}$ is the best-fit photometric redshift estimate as reported by the Directional Neighborhood Fitting (DNF) algorithm~\citep{DeVicente2016}, and $a=4.0$, $b=18$. The sample was constructed in \citet{y3-2x2maglimforecast} to optimize cosmological constraints obtained from galaxy clustering and galaxy-galaxy lensing and is split into 6 tomographic bins. We use the redshift distribution for each tomographic bin as estimated by~\citet{Porredon2021}.
These distributions are allowed to shift and stretch with priors from~\citet{2020arXiv201212826C}. The redshift distributions have been further validated using a self-organizing map method~\citep{y3-2x2ptaltlenssompz}. Normalized redshift distributions are shown in Figure~\ref{fig:zdists}, and galaxy number counts per redshift bin are shown in Table~\ref{tab:samples}. In order to correct for the impact of survey properties on galaxy number density, each galaxy acquires a weight obtained by~\citet{Rodriguez-Monroy2021}. In \citet{DES2021}, only the first four bins were used in the fiducial cosmology analysis since it was found that the model was not able to give a good fit to the two highest redshift bins. As a result, in this work we also only use the first four bins for the fiducial analysis. However, the results in~\citet{y3-5x2-ii} show that the 2-point clustering measurements in these high redshift bins are consistent with the combination of galaxy clustering+CMB-lensing and galaxy shear+CMB-lensing. This suggests that the issues found for these high-redshift bins in~\citet{DES2021} are more likely due to the galaxy-galaxy lensing measurements. We will, thus, also show how our results could change when including the high-redshift bins.

\begin{table}
    \centering
    \begin{tabular}{c|c}
    \multicolumn{2}{c}{\maglim{}}\\
    \hline
    Redshift bin & $N_{\rm gal}$ \\
    \hline
    $0.20 < z < 0.40$ & 2236462\\
    $0.40 < z < 0.55$ & 1599487\\
    $0.55 < z < 0.70$ & 1627408\\
    $0.70 < z < 0.85$ & 2175171\\
    $0.85 < z < 0.95$ & 1583679\\
    $0.95 < z < 1.05$ & 1494243\\
    \hline
    \end{tabular}
    \caption{Redshift slices and number of galaxies used in this study for the \maglim sample.}
    \label{tab:samples}
\end{table}

\subsection{$y$-maps} 

\begin{figure*}
    \centering
    \includegraphics[width=0.6\textwidth, trim={2cm 4cm 3cm 4cm},clip]{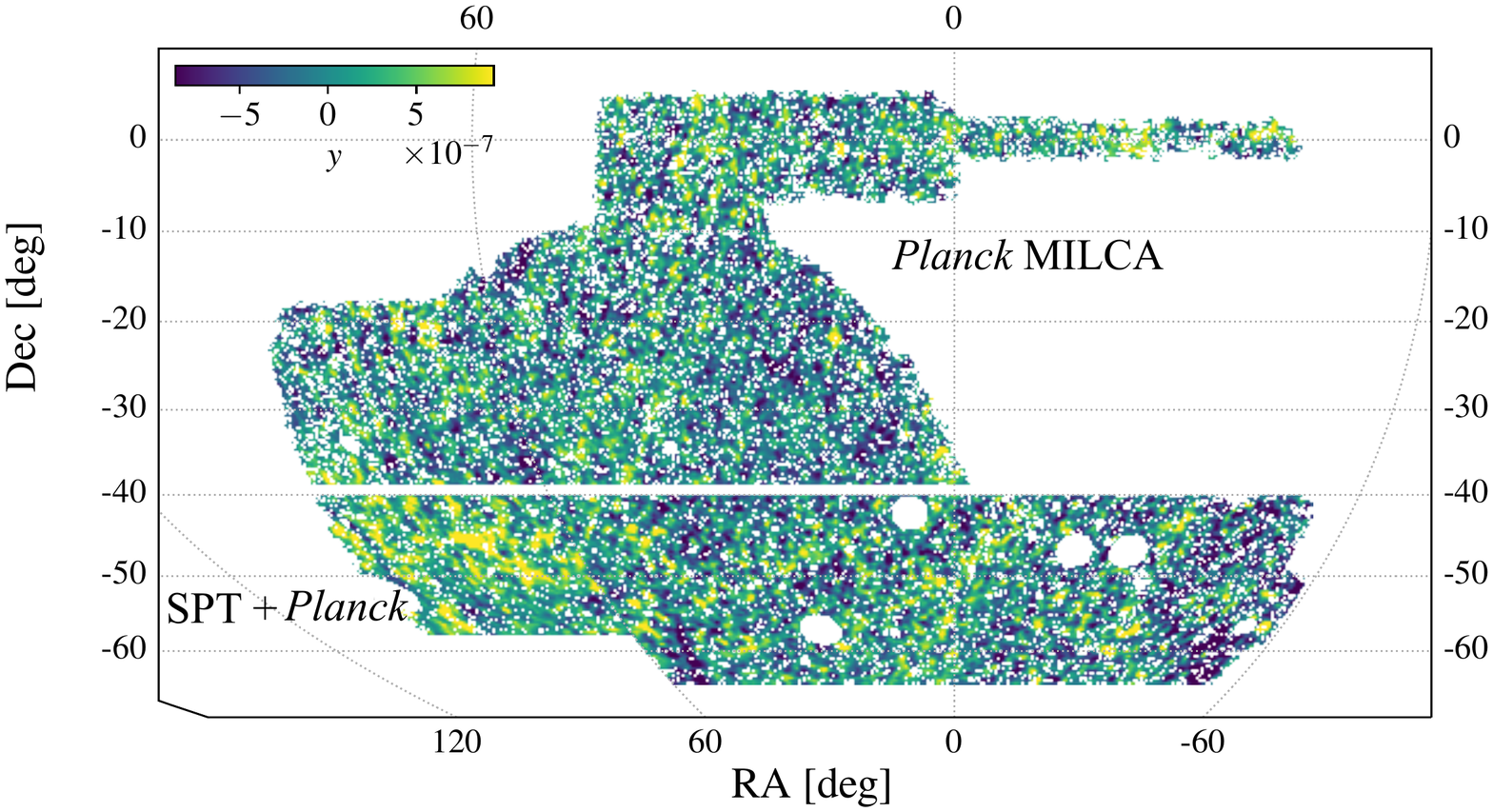}
    \caption{Combined SPT+{\it Planck} $y$-map in equatorial coordinates displayed using an equal-area McBryde-Thomas flat-polar quartic projection. The southern region (${\rm Dec} < -40$ deg.) corresponds to the SPT+{\it Planck} minimum variance $y$-map from \citet{Bleem2021}, and the northern region (${\rm Dec} > -39$ deg.) corresponds to the MILCA $y$-map from~\citet{Planck2016}. We choose to have a gap between the two regions in order to improve the level of independence of our results. We transform the original data products to \textsc{healpix} resolution $N_{\rm{side}}=2048$. For display purposes, the maps shown have been smoothed with a Gaussian (FWHM$=0.25$ deg.) beam.}
    \label{fig:ymap}
\end{figure*}

As mentioned previously, the tSZ effect is typically measured in terms of the Compton-$y$ parameter. Typically, $y$-maps are built using a linear combination of individual frequency mm-wave/microwave maps~\citep[see][for a review]{Delabrouille&Cardoso2009} and contain valuable cosmological information~\citep[][and references therein]{Komatsu&Seljak2002}. In this work however, we focus on the cross-correlation between $y$ and the galaxy counts. The southernmost part of the footprint of DES was designed to overlap with the South Pole Telescope's SPT-SZ survey area~\citep{Story2013}. However, there is a significant fraction of the DES Y3 footprint that does not overlap with the SPT-SZ survey. Therefore, we use the SPT-SZ + \Planck maps for Dec $< -40$ deg. and the \Planck MILCA $y$-map~\citep{Planck2016} for Dec $> -39$ deg. (which we will refer to as MILCA hereafter). Both of the $y$-maps that we use in our analysis are shown in Figure~\ref{fig:ymap}. We provide additional details about the maps we use in the following subsections.

\subsubsection{SPT-SZ+{\it Planck} $y$-maps from \citet{Bleem2021} }

For this work we focus on the component-separated $y$-maps using a combination of data from the SPT-SZ survey and {\it Planck} \citep{Bleem2021} which are publicly available.\footnote{\url{https://lambda.gsfc.nasa.gov/product/spt/spt_prod_table.cfm}} The maps cover $\sim 2500$ square degrees of the southern sky (with $\sim 1800$ square degrees overlapping with DES Y3 galaxies) with a 1.25 arcmin resolution. In this work, we use the minimum variance $y$-map presented in \citet{Bleem2021} for our fiducial measurements because it has the lowest noise and the smallest beam size. Additionally, we use the CMB-CIB-nulled $y$-map to test the presence of CIB contamination in our measurements. For details about these maps, the algorithms behind their construction, and their validation, we refer the readers to~\citet{Bleem2021}.  

In addition to the publicly available maps, we test a custom CIB-reduced (which we will refer to as CIB-nulled) map to ensure a low-level of CIB contamination. This map is generated using the same $y$-map implementation presented in \citet{Bleem2021}. The main difference between the CMB-CIB-nulled (or ``three-component") map from \citet{Bleem2021} and our CIB-nulled map (which would be a ``two-component" map), is that for the latter we focus on minimizing the residual CIB using the CIB model presented in~\citet{Reichardt2021}, whereas the CMB-CIB-nulled map not only tries to minimize the CIB residual but also the CMB (which should not correlate with DES galaxies). This results in the CMB-CIB-nulled map having a higher noise level.

\subsubsection{MILCA $y$-map from \citet{Planck2016}}

We follow previous analyses~\citep{Planck2016, Hurier&Lacasa17, Pandey2019, Koukoufilippas2020} and use the MILCA $y$-map from the \textit{Planck} collaboration~\citep{Planck2016}. This map\footnote{The maps and masks used are publicly available at \url{https://irsa.ipac.caltech.edu/data/Planck/release_2/all-sky-maps/ysz_index.html}.} has a beam size of FWHM=10 arcmin. We apply the 40\% Galactic mask and point source mask presented in \citet{Planck2016}. The MILCA $y$-map is generated by combining different frequency maps from the \textit{Planck} mission. This reconstruction method allows the introduction of external templates to remove unwanted components, such as the CIB. However, the minimization of the CIB signal depends on the particular templates used, and, as the CIB-induced bias decreases, the noise level increases. Moreover, as the CIB maps are not totally correlated across frequencies, its contribution cannot be fully removed~\citep{Hurier2013}. 
Therefore, although the MILCA $y$-map uses CIB templates to minimize the CIB contamination, the effect of any residual CIB leakage on the cross-correlation with galaxies or shear should still be considered carefully.

\section{Analysis}
\label{sec:analysis}

In this section we describe the different components of the analysis. In Section~\ref{sec:measurement} we describe how we construct the data vector; in Section~\ref{sec:cov} we describe the covariance matrix we use; in Section~\ref{ssec:scale_cuts} we describe our choice of scale cuts; in Section~\ref{sec:likelihood} we introduce our inference framework; in Section~\ref{ssec:systematics} we perform a series of diagnostic tests on our measurements to ensure there is no significant systematic contamination.

\begin{figure*}
    \centering
    \includegraphics[width=0.83\textwidth]{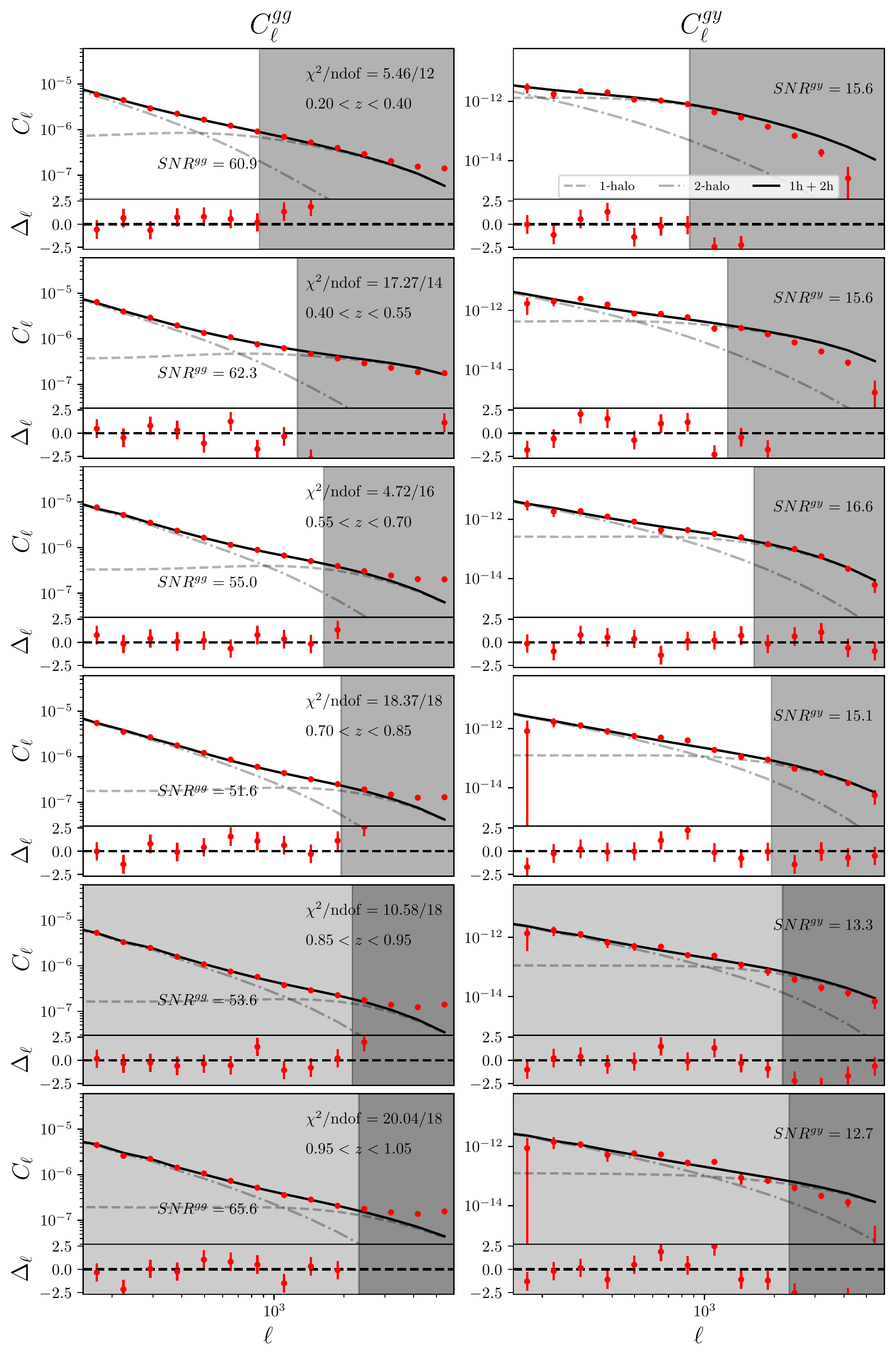}
    \caption{Measured galaxy-galaxy (left column, $C^{\rm gg}_{\ell}$) and galaxy-$y$ (right column, $C_{\ell}^{\rm gy}$) power spectra and best fit HOD (solid black lines) for the southern of the DES footprint using the SPT-SZ+\Planck $y$-map. The lower sub-panels show the residuals divided by the uncertainty, which we denote as $\Delta_{\ell}$. The signal-to-noise ratio are annotated for each redshift bin. The $\chi^{2}/\rm{ndof}$ values included in the left panels correspond to the total ($\rm gg + gy)$ $\chi^2$.}
    \label{fig:data_vector_spt}
\end{figure*}

\begin{figure*}
    \centering
    \includegraphics[width=0.83\textwidth]{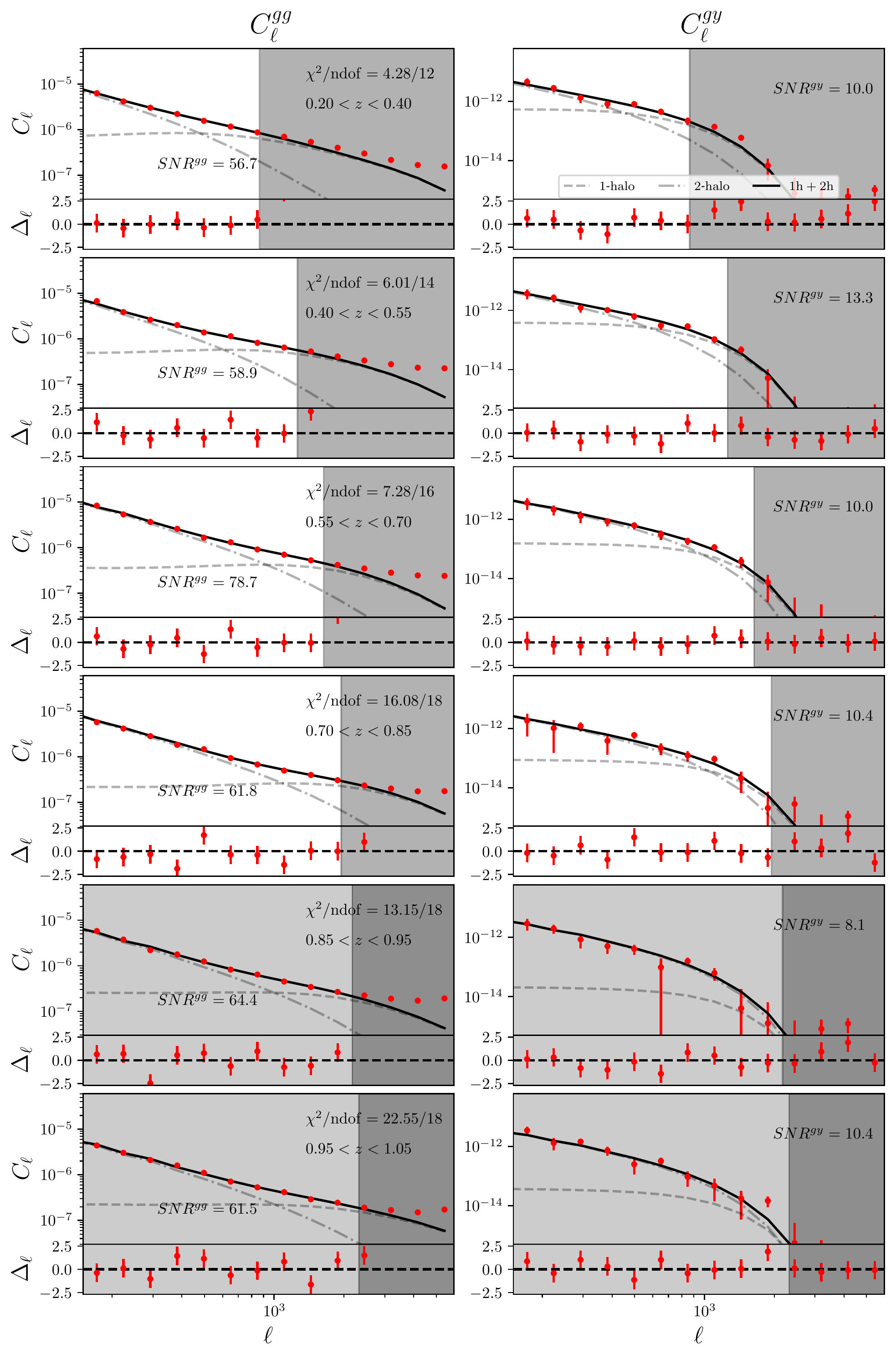}
    \caption{Measured galaxy-galaxy (left column, $C^{\rm gg}_{\ell}$) and galaxy-$y$ (right column, $C_{\ell}^{\rm gy}$) power spectra and best-fit HOD (solid black lines) for the northern region of the DES footprint using the MILCA $y$-map correcting for the CIB contribution (see details in Section~\ref{ssec:systematics}). The lower subpanels show the residuals divided by the uncertainty, which we denote as $\Delta_{\ell}$. More details are described in Figure~\ref{fig:data_vector_spt}.}
    \label{fig:data_vector_planck}
\end{figure*}

\subsection{Measurement}
\label{sec:measurement}
 
Using the galaxy samples and $y$-maps described in the previous sections as input, we measure the power spectrum using the pseudo-$C_{\ell}$ \texttt{MASTER} algorithm~\citep{Hivon2002} as implemented in \texttt{NaMaster}~\citep{NaMaster}. To do this, we first construct a galaxy density map from the galaxy catalog by filling each pixel of the map by $\delta_{g}=\frac{n_{g} - \bar{n}_{g}}{\bar{n}_{g}}$, where $n_{g}$ is the number of galaxies in each pixel and $\bar{n}_{g}$ is the mean galaxy number count per pixel. This is done for each of the tomographic bins and the map is constructed with a \textsc{healpix} format of $N_{\rm side}=2048$. The $y$-maps are downgraded to $N_{\rm side}=2048$ to match the galaxy density maps. Alongside the galaxy density maps, we generate weight maps by summing the weights of each galaxy within a pixel. The weights here are designed to correct for the effect of systematic effects that are imprinted via different survey properties described in~\citet{Rodriguez-Monroy2021}. 

We use logarithmic binning in $\ell$ with the following bandpower edges: $\left[ \right.$150, 195, 254, 332, 433, 564, 736, 959, 1251, 1631, 2126, 2772, 3614, 4712, 6142$\left.\right]$. The measured power spectra are shown in Figures~\ref{fig:data_vector_spt} and~\ref{fig:data_vector_planck} for the southern (using the SPT-SZ+\Planck maps) and northern (using the MILCA map) regions, respectively. The grey shaded areas are not considered in our fiducial analysis (see Section~\ref{ssec:scale_cuts}). Comparing Figure~\ref{fig:data_vector_spt} and~\ref{fig:data_vector_planck}, the measured $C^{\rm gg}_{\ell}$ agree well between the two regions of sky at the scales we consider. This is a good first check to show that the galaxy sample is homogeneous across the sky. 
For $C^{\rm gy}_{\ell}$, while the large scale measurements are in good agreement, we notice some differences at small scales ($\ell > 500$) between the two regions. This is a consequence of the different smoothing scale used in the SPT-SZ+\Planck and MILCA $y$-maps.

In Figures~\ref{fig:data_vector_spt} and~\ref{fig:data_vector_planck} we also quote the signal-to-noise ratio (SNR), which we calculate as $\mathrm{SNR}^{uv} = \sqrt{C_{\ell}^{uv} \mathcal{C}_{\ell,\ell'}^{-1}C_{\ell'}^{uv}}$, where $\mathcal{C}_{\ell, \ell'}^{-1}$ is the inverse covariance matrix. We restrict the calculation of the SNR to the scales that are considered in our fiducial analysis $\ell > 150$ and $k < 0.7$ Mpc$^{-1}$. The choice of these scale-cuts is discussed in Section~\ref{ssec:scale_cuts}. 

\subsection{Covariance matrix}
\label{sec:cov}

We use a jackknife (JK) covariance matrix in this analysis in order to appropriately capture any spatial variation in the data beyond the analytical model. We use 75 and 92 roughly equal-area JK patches in the southern and northern regions. 
The JK patches are generated via the $k$-means algorithm implementation included in the \texttt{TreeCorr} package \citep{TreeCorr}.

The JK-estimated mean data vector is 
\begin{equation} 
    \langle X (\ell) \rangle = \frac{1}{N_{\rm JK} - 1}\sum_{i}^{N_{\rm JK}} X_{i} (\ell), 
\end{equation}
where $X$ corresponds to $C^{\rm gg}$ or $C^{\rm gy}$ here, $N_{\rm JK}$ is the number of JK patches, $i$ indicates individual measurements of $X(\ell)$ leaving one JK patch $i$ out. The corresponding covariance matrix $\mathcal{C}$ is given by
\begin{equation} 
    \mathcal{C} = \frac{N_{\rm JK} -1}{N_{\rm JK}}\sum (X_i - \langle X (\ell) \rangle)^{T} (X_i - \langle X (\ell) \rangle). 
\end{equation}
 We compare our JK estimates with the analytical covariance computed as described in~\citet{Koukoufilippas2020}, finding agreement within $20\%$ for all bins (where the JK-estimated covariance is larger). 

As the jackknife covariance is known to be noisy and introduces a bias when inverting, we follow \citet{kaufman1967, hartlap2007} 
and multiply it by a factor $H$ to get the unbiased covariance
\begin{equation}\label{eq:HartlapCocariance}
	\mathcal{C}_H^{-1} = H \mathcal{C}^{-1} = \left(\frac{N_{\rm JK} - N_{\rm band} - 2}{N_{\rm JK}-1}\right) \mathcal{C}^{-1} \; ,
\end{equation}
where $N_{\rm band}$ is the number of bandpowers we use.



\subsection{Scale cuts}
\label{ssec:scale_cuts}

In this work we explore two regimes of the data vectors separately: the large, linear scales and the small, highly nonlinear scales. The fact that we have not been able to coherently model the two regimes under the same model is mainly limited by our ability to model the 1-to-2 halo transition region for $C_{\ell}^{\rm gg}$, which has been known to be challenging \citep{Mead15, Hadzhiyska20}. 
This motivates us first to look separately at the large-scale results, which are more robust to the small-scale modelling uncertainties, and then explore ways to extract information on smaller scales with some assumptions on the HOD. The small-scale analysis also takes advantage of the SPT-SZ + \Planck $y$-maps which are higher resolution than the MILCA $y$-map, allowing us to probe further into the 1-halo term of $C_{\ell}^{\rm gy}$. 

For the large-scale analysis, we use scales $\ell < \ell_{\rm max}$, with $\ell_{\rm max, i} = k_{\rm max} \chi(\bar{z}_{i}) - 1/2$, with $k_{\rm max} = 0.7$ Mpc$^{-1}$ (which corresponds to $\ell_{\rm max} = 864, 1259, 1636, 1946, 2171, 2320$ for redshift bins 0 to 5, respectively), and $\chi(\bar{z}_{i})$ the comoving distance at the mean redshift of each bin. 
In addition, we ignore the modes below $\ell_{\rm min} = 150$ since our jackknife covariances are not accurate for these modes, given that the typical jackknife region is smaller than the modes that we want to map. For our large-scale analysis, we apply these same cuts to $C_{\ell}^{\rm gy}$. The priors associated with the model parameters for the large-scale analysis are listed in Table~\ref{tab:priors_large_scales}.

For the small-scale analysis (see Section~\ref{sec:results2}), we rely on the HOD and bias parameters obtained from the large-scale results, and fit our model to the measured $C_{\ell}^{gy}$ using the parameters and priors described in Table~\ref{tab:priors_small_scales} for the tSZ profile in Equation~\ref{eq:gas_prof}. We fix $\gamma=0.31$ as in~\citet{2010A&A...517A..92A}, as we notice that we are  insensitive to the value of this parameter. For our analysis in this regime, we restrict our analysis to scales larger than $k_{\rm max} = 2.5$ Mpc$^{-1}$, as we observed that the modeling starts to fail to describe smaller scales. Furthermore, as shown in \citet{Rodriguez-Monroy2021}, there is indication that the correction of systematic effects in the galaxy clustering measurement is less effective on the smallest scales. These effects were included at the covariance level for~\citet{DES2021, Porredon2021} in real space.

\begin{table}
    \centering
    \begin{tabular}{ccc}
        Parameter & Fiducial & Prior \\
        \hline
        $\log_{10} M_{\rm min}/M_{\odot}$ & & $\mathcal{U}(10, 16)$\\
        $\sigma_{\rm{ln}M}$ & 0.15 & Fixed \\
        $\log_{10} M_{0}/M_{\odot}$ & $\log_{10} M_{\rm min}/M_{\odot}$ & \\
        $\log_{10} M'_{1}/M_{\odot}$ & & $\mathcal{U}(10, 16)$ \\
        $\alpha_{s}$ & & $\mathcal{U}(0, 3)$ \\
        $f_{c}$ & $1$ & Fixed\\
        $\beta_{g}$ & & $\mathcal{U}(0.1, 10)$\\
        $\beta_{\rm max}$ & & $\mathcal{U}(0.1, 10)$\\
        $\rho_{\rm gy}$ & & $\mathcal{U}(-1, 1)$\\
        $b_{H}$ & & $\mathcal{U}(0, 1)$\\
        \hline
        $\sigma_{z, 0}$ & 0.975 & $\mathcal{N}(0.975, 0.062)$\\
        $\Delta z_{0}$ & -0.009 & $\mathcal{N}(-0.009, 0.007)$\\
        $\sigma_{z, 1}$ & 1.306 & $\mathcal{N}(1.306, 0.093)$\\
        $\Delta z_{1}$ & -0.035 & $\mathcal{N}(-0.035, 0.01)$\\
        $\sigma_{z, 2}$ & 0.870 & $\mathcal{N}(0.870, 0.054)$\\
        $\Delta z_{2}$ & -0.005 & $\mathcal{N}(-0.005, 0.006)$\\
        $\sigma_{z, 3}$ & 0.918 & $\mathcal{N}(0.918, 0.051)$\\
        $\Delta z_{3}$ & -0.007 & $\mathcal{N}(-0.007, 0.006)$\\
        $\sigma_{z, 4}$ & 1.080 & $\mathcal{N}(1.080, 0.067)$\\
        $\Delta z_{4}$ & 0.002 & $\mathcal{N}(0.002, 0.007)$\\
        $\sigma_{z, 5}$ & 0.845 & $\mathcal{N}(0.845, 0.073)$\\
        $\Delta z_{5}$ & 0.002 & $\mathcal{N}(0.002, 0.008)$\\
        \hline
    \end{tabular}
    \caption{List of parameters and priors used for the large-scale hydrostatic bias results. We follow \citet{Ando2018, Koukoufilippas2020} and set $M_{\rm min}=M_{0}$. For the photo-$z$ nuisance parameters $\Delta z_{i}, \sigma_{z, i}$ we use the parametrizations and priors suggested by~\citet{2020arXiv201212826C}.}
    \label{tab:priors_large_scales}
\end{table}

\begin{table}
    \centering
    \begin{tabular}{c|c}
        Parameter &  Prior\\
        \hline
        $\alpha$ & $\mathcal{U}(0.3, 4)$\\
        $\beta$ & $\mathcal{U}(0.2, 10)$\\
        $c_{P}$ & $\mathcal{U}(0.1, 5)$\\
        $\gamma$ & 0.31\\
        \hline
    \end{tabular}
    \caption{Parameters and priors used in the small-scale analysis. We fix $\gamma$ to the value found in~\citet{2010A&A...517A..92A}.}
    \label{tab:priors_small_scales}
\end{table}


\subsection{Likelihood and inference}
\label{sec:likelihood}


 We assume a Gaussian likelihood for the data vector of measured correlation functions, $\vec{d}$, given a model, $\vec{m}$, generated using the set of parameters $\vec{p}$:
\begin{multline}
\ln \mathcal{L}(\vec{d}|\vec{m}(\vec{p}))=  
-\frac{1}{2} \sum^N_{ij} \left(d_i - m_i(\vec{p})\right)^T \mathbf{C}^{-1}_{ij} \left(d_j - m_j(\vec{p}) \right),  
\end{multline}
where the sums run over all of the $N$ elements in the data and model vectors. The posterior on the model parameters is then given by:
\begin{equation}
P(\vec{m}(\vec{p})|\vec{d}) \propto \mathcal{L}(\vec{d} | \vec{m}(\vec{p})) P_{\rm prior} (\vec{p}),
\end{equation}
where $P_{\rm prior}(\vec{p})$ is a prior on the model parameters. 

We sample the posterior by running a Markov Chain Monte Carlo using the ensemble sampler implemented in \texttt{emcee}~\citep{emcee}.

\subsection{Systematics tests}
\label{ssec:systematics}
Precision measurements of the galaxy power spectrum $C_{\ell}^{\rm gg}$ can be affected by observing conditions or the presence of bright objects, which can lead to systematic biases in the measured $C_{\ell}^{\rm gg}$. In other studies using these samples~\citep{DES2021, Rodriguez-Monroy2021, Porredon2021, Pandey2021} these systematic biases are mitigated by the usage of weights on each galaxy \cite[for details see][]{Rodriguez-Monroy2021}. We follow the same approach and adopt the weights as described in Section~\ref{sec:measurement}. The galaxy-$y$ cross-correlation $C_{\ell}^{\rm gy}$, on the other hand, should be much less affected by the particular observing conditions of the DES galaxies since it is a cross-correlation measurement of two independent datasets. However, $C_{\ell}^{\rm gy}$ can be affected by other astrophysical systematic effects associated with foregrounds. We test the effect of three of them: dust, bright radio sources, and the cosmic infrared background (CIB). 

\begin{itemize}
    \item \textbf{Dust:} We use two sets of independent reddening maps, the maps presented in~\citet{Delchambre22} and the maps presented in~\citet{Lenz2017}, and compare the power spectra results with and without deprojecting~\citep{Elsner2017, NaMaster} these maps from the galaxy and $y$-maps. The idea behind this test is the following: deprojection (or mode projection) essentially assumes that the true signal of interest (in our case $C^{gy}_{\ell}$) is not correlated with the foregrounds, and uses a template of the foregrounds to obtain the cleaned, unbiased power spectra. In the case that the foregrounds are correlated with the signal, deprojection can produce over-corrected power spectra. This means, if the difference between the deprojected signal and the non-deprojected signal is not statistically significant, the impact of the foreground in the signal is likely small. We carry out the full analysis with deprojected and non-deprojected galaxy auto spectrum and galaxy-$y$ cross spectrum. We find that the maximum absolute value across all redshift bins for the shift of the best-fit $(1-b_{H})$ is $\approx 0.23\sigma$, where $\sigma$ is the statistical uncertainty.\\
    
    
    \item \textbf{Bright radio sources:} Radio sources are a known contaminant of the tSZ maps~\citep{Bleem2021}. For this test we focus on the MILCA map, as their masking threshold is substantially larger than the SPT-SZ + {\it Planck} maps (${\sim}200$ mJy at {\it Planck} 143 GHz which goes into MILCA vs $\sim$6 mJy at 150 GHz for SPT-SZ). We check the effect of these sources by applying an additional mask for 
    sources brighter than 1 mJy in the 1.4 GHz NVSS catalog~\citep{NVSS98}. We find that the maximum absolute value for the shift across all redshift bins of the best-fit $(1-b_{H})$ is $\approx 0.15\sigma$ when this additional mask is applied.\\
    
    
    \item \textbf{CIB:} One of the most important contaminants of $C_{\ell}^{\rm gy}$ is the CIB, as it is correlated with both the galaxy positions and the $y$-map~\citep{Koukoufilippas2020, Chiang2020, Gatti2021}. In order to test for the effect of CIB contamination in our measurements we estimate the CIB leakage in $C_{\ell}^{\rm gy}$. The idea is the following: 
    The SPT-SZ+\textit{Planck} $y$-maps and the MILCA map are constructed using a linear combination of frequency channels. Schematically this can be written as:
    \begin{equation}
    y^{\rm ILC}(p)={\bf w}^{t}x(p),
    \end{equation}
    where $y^{\rm ILC}$ denotes the component separated $y$-map, ${\bf w}^{t}$ is a vector containing weights (or coefficients) to be multiplied to the individual frequency channels\footnote{The weights for the SPT map can be found at \url{https://pole.uchicago.edu/public/data/sptsz_ymap/component_map_weights/}, while the weights for the MILCA map can be found at \url{https://wiki.cosmos.esa.int/planck-legacy-archive/images/8/83/Milca_nilc_IDL_routines.zip}. The MILCA weights used here are also spatially varying and therefore ${\bf w}^{t}$ is a function of direction. }, and $x(p)$ is a vector containing the individual frequency channels.  In reality, the frequency channel maps contain various foregrounds: 
    \begin{align}
    x(p,\nu)&=x^{\rm CMB}(p,\nu)+x^{\rm CIB}(p,\nu)\nonumber\\
    &+x^{\rm tSZ}(p,\nu)+x^{\rm kSZ}(p,\nu)+x^{\rm radio}(p,\nu)+ \ldots
    \end{align}
    Since the relationship is linear, we can see that the amount of CIB residuals in our map can be computed using:
    \begin{equation}
    y^{\rm CIB}(p)={\bf w}^{t}x^{\rm CIB}(p,\nu).
    \end{equation}
    This implies that we need a clean map of the CIB for every frequency channel that goes into the component separation algorithm. Unfortunately, while we do have relatively clean maps of the CIB at higher frequencies (353/545/857 GHz), such maps do not exist for the lower frequency channels ({\it Planck} 100/143/217 or SPT 95/150/220 GHz) since it is challenging to disentangle CIB from other astrophysical components at the those frequencies. Therefore, we make predictions of these maps by taking the 353 GHz map from \cite{Lenz2019}, and scaling down the amplitude of that map assuming an SED of the CIB.\footnote{For this, we use the best-fit CIB power spectra templates from \cite{Reichardt2021} (and also extrapolations to 353 GHz). We compute the mean amplitude of the CIB model in the $\ell$ range $100<\ell<1500$ for all the frequency channels, and take the scaling factors as the ratio between those values and the value for 353 GHz.}
    
    Using this convention, we estimate the CIB residual in the $y$-maps, $y^{\rm CIB}$, in each of the maps used in this analysis. We then compute the cross-correlation between each $y^{\rm{CIB}}$-maps and our galaxy samples, $C_{\ell}^{gy,{\rm CIB}}$, and estimate the ratio between this measurement and the original power spectra. We find that, in the range of scales considered, the median value of this ratio is $< 1\%$ for the minimum variance SPT-SZ + \Planck map, as shown in Figure~\ref{fig:ycib_x_gal}. This is not the case for the MILCA $y$-map, which show significant levels of CIB contamination for the last 3 bins considered in our analysis, in line with the results found by~\citet{Pandey2019, Gatti2021}. 
    
    \begin{figure*}
    \centering
    \includegraphics[width=0.98\textwidth]{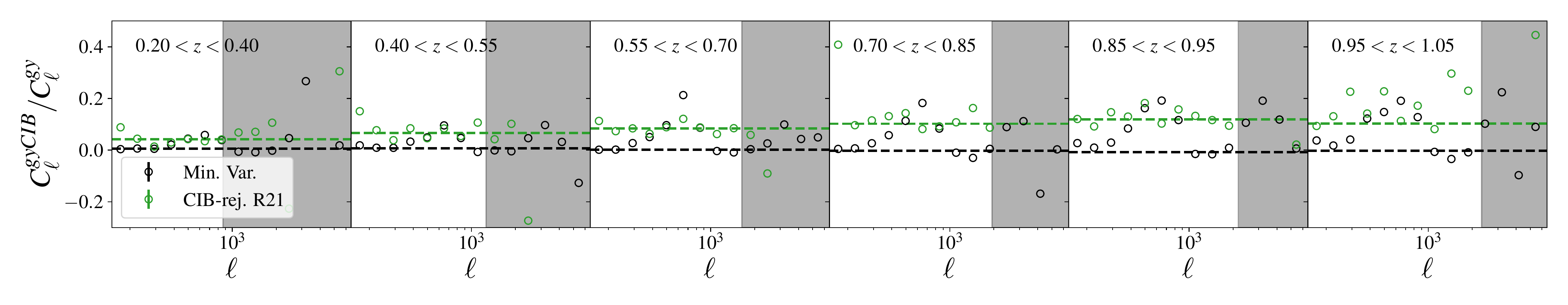}
     \includegraphics[width=0.98\textwidth]{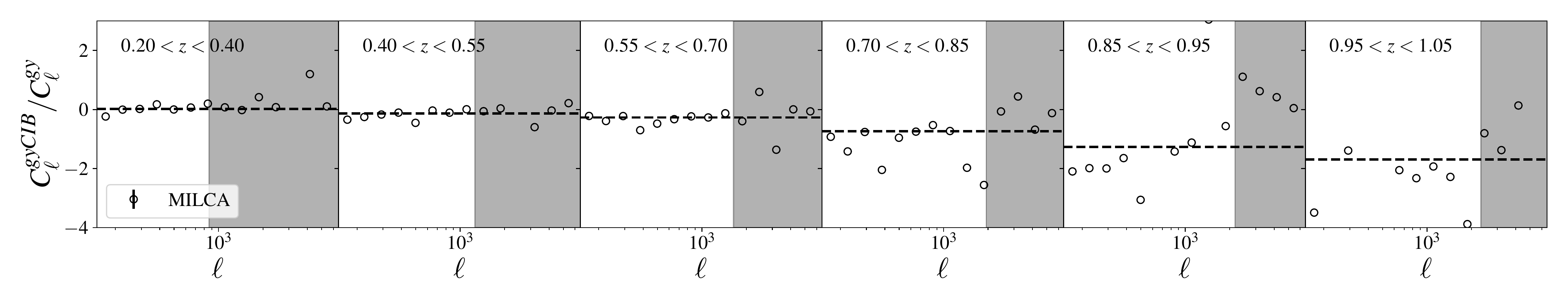}
    \caption{Top panel: Relative CIB leakage in the SPT-SZ + {\it Planck} maps, $C_{\ell}^{gy^{CIB}}/C_{\ell}^{yg}$. The dashed line represents the median ratio within the considered scale cuts. Bottom Panel: CIB leakage for MILCA $y$-map. For the last 2 redshift bins the CIB signal is dominant.}
    \label{fig:ycib_x_gal}
    \end{figure*}
    
    As a second robustness test, we compare our fiducial measurements using the minimum variance map from~\citet{Bleem2021} with the measurements using CIB-CMB-nulled $y$-map from \citet{Bleem2021}. In addition to this map, we also compare with the custom CIB-nulled version of the $y$-map described in Section~\ref{sec:data_sample}. In Figure~\ref{fig:bH_comparison} we show the best-fit $\left(1-b_{H}\right)$ values for different SPT-SZ + \Planck maps. For the MILCA map, we show the results for the original maps, and after correcting for CIB. This correction consists on subtracting the contribution of $y^{\rm{CIB}}$ described above from the original $y$-map, i.e, $y^{\rm MILCA-corr} = y^{\rm MILCA} - y^{\rm CIB}$. We see that the different SPT-SZ + {\it Planck} results are compatible with each other, and compatible with the MILCA  results after applying the CIB correction. Given the size of the systematic CIB correction using the MILCA map, we choose not to combine the CIB-corrected MILCA $(1-b_{H})$ measurements with those obtained from SPT-SZ + {\it Planck} in our final results, and only use it as a cross-check here.
\end{itemize}

\begin{figure}
    \centering
    \includegraphics[width=0.45\textwidth]{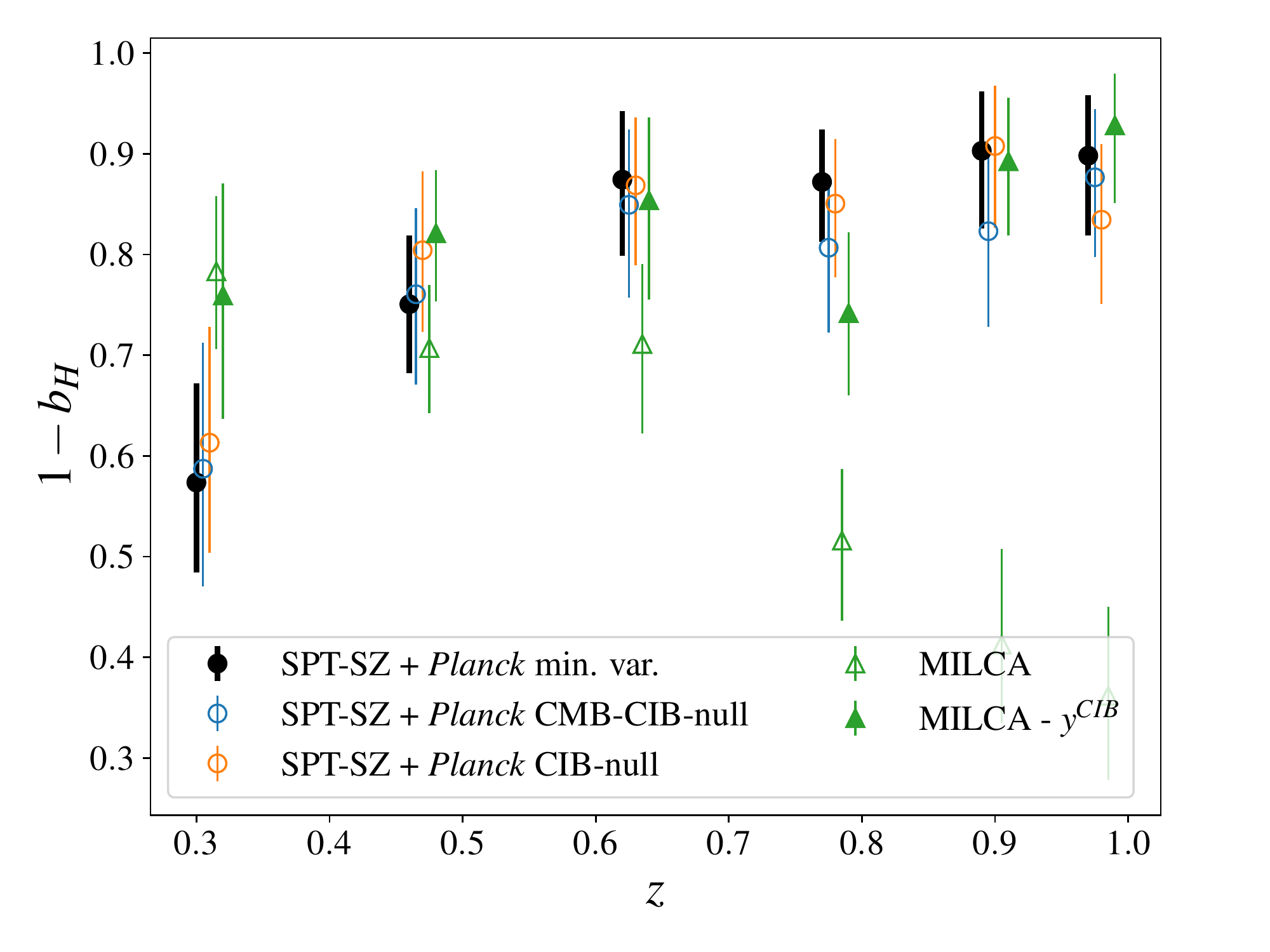}
    \caption{Best-fit $\left(1-b_{H}\right)$ values as a function of redshift for different maps, showing the impact of CIB on these measurements. Our fiducial values (SPT-SZ + \Planck min. var.) are shown as the solid black circles.}
    \label{fig:bH_comparison}
\end{figure}

\section{Large-scale analysis: constraints on hydrostatic mass bias}
\label{sec:results1}


In this section we examine the large-scale constraints from the $C_{\ell}^{\rm gg}+C_{\ell}^{\rm gy}$ measurements. As discussed in Section~\ref{sec:intro}, on these large scales, since $C_{\ell}^{\rm gg}\propto b_{g}^{2}$, where $b_{g}$ is the galaxy bias, and $C_{\ell}^{\rm gy}\propto b_{g} \langle b_{h}P_{e}\rangle $, the combination helps us constrain the quantity $\langle b_{h}P_{e}\rangle$, which is directly related to the hydrostatic mass bias $b_{H}$.


We use the modeling framework described in Section~\ref{sec:methods} and the methodology described in Section~\ref{sec:analysis}, fitting the large scales with $k \leq 0.7$ Mpc$^{-1}$. In order to make a fair comparison with other studies in the literature, we  fix the gas profile parameters to their fiducial values $(\alpha, \beta, \gamma, c_{P}) = (1.33, 4.13, 0.31, 1.81)$~\citep{Planck2013a}. We derive the model fits shown in Figures~\ref{fig:data_vector_spt} and~\ref{fig:data_vector_planck}. The grey dash (dotted-dash) lines show the 1-halo (2-halo) component of the fit and the solid black line shows the full model. In each panel the lower sub-panel shows the residual of the fit divided by the uncertainty. In general, we find good fits to the measurements in all redshift bins, including the last two. The goodness-of-fit and corresponding PTE are shown in Table~\ref{tab:gy_SNR_bPe}.  The good quality of the fits in the last two bins is in agreement with the interpretation put forward by \citet{y3-5x2-ii}, that there appears to not be significant systematic contamination in the galaxy clustering measurements for these bins. One can also see clearly the $C_{\ell}^{\rm gg}$ data points deviating from the model at the smallest scales (large $\ell$) which we do not use for the fits -- there is excess power in the measurements that is not described by a typical 1-halo term. This deviation starts fairly consistently at $\ell \sim 3000$ (corresponding to $\theta \sim 3.6$ arcmin) across all bins and is consistent with what was found in \citet{Rodriguez-Monroy2021}, where the small-scale galaxy clustering could be contaminated by systematic effects (see Figure 6 in that paper). 

\begin{table}
    \centering
    \begin{tabular}{cccccc}
    Bin & $\langle z \rangle$ & $\chi^{2}/$ndof & PTE & $b_{H}$ & $\langle b_{h}P_{e} \rangle$ [meV cm$^{-3}$] \\
    \hline
\multicolumn{6}{l}{\textsc{SPT-SZ + Planck}} \\
0 & $0.30$ & 5.46/12 & 0.94 & 
    $0.43^{+0.09}_{-0.10}$ & $0.16 ^{+0.03}_{-0.04}$   \\
1 & $0.46$ & 17.27/14 & 0.24 & 
    $0.25^{+0.07}_{-0.07}$ & $0.28 ^{+0.04}_{-0.05}$   \\
2 & $0.62$ & 4.72/16 & 1.00 & 
    $0.13^{+0.08}_{-0.07}$ & $0.45 ^{+0.06}_{-0.10}$   \\
3 & $0.77$ & 18.37/18 & 0.43 & 
    $0.13^{+0.06}_{-0.05}$ & $0.54 ^{+0.08}_{-0.07}$   \\
\hline
\g{4} & \g{$0.89$} & \g{10.58/18} & \g{0.91} & 
    \g{$0.10^{+0.08}_{-0.06}$} & \g{$0.61 ^{+0.08}_{-0.06}$}   \\
\g{5} & \g{$0.97$} & \g{20.04/18} & \g{0.33} & 
    \g{$0.10^{+0.08}_{-0.06}$} & \g{$0.63 ^{+0.07}_{-0.08}$}   \\
\hline
\multicolumn{6}{l}{\textsc{MILCA} (CIB corrected)} \\
0 & $0.30$ & 4.28/12 & 0.99  & $0.24 ^{+0.12}_{-0.11}$  &  $0.24 ^{+0.09}_{-0.07}$ \\
1 & $0.46$ &  6.01/14 & 0.99 & $0.19 ^{+0.07}_{-0.06}$ &   $ 0.32 ^{+0.04}_{-0.06}$\\
2 & $0.62$ & 7.28/16 & 0.99 & $0.15 ^{+0.10}_{-0.08}$ & $0.51 ^{+0.17}_{-0.08}$  \\
3 & $0.77$ & 16.08/18 & 0.62 & $0.26 ^{+0.08}_{-0.08}$  & $0.40^{+0.08}_{-0.11}$  \\
\hline
\g{4} & \g{$0.89$} & \g{13.15/18} & \g{0.86}  & \g{$0.11 ^{+0.07}_{-0.06}$} & \g{$0.57 ^{+0.07}_{-0.07}$}  \\
\g{5} & \g{$0.97$} & \g{22.55/18} & \g{0.14}  & \g{$0.07 ^{+0.08}_{-0.05}$}  & \g{$0.65 ^{+0.08}_{-0.10}$}\\
\hline
\multicolumn{6}{l}{\textsc{MILCA} (raw)} \\
0 & $0.30$ & 3.17/12 & 1.00  & $0.22 ^{+0.08}_{-0.08}$  &  $0.27 ^{+0.05}_{-0.05}$ \\
1 & $0.46$ &  6.97/14 & 0.99 & $0.29 ^{+0.06}_{-0.06}$ &   $ 0.28 ^{+0.04}_{-0.04}$\\
2 & $0.62$ & 13.13/16 & 0.73 & $0.29 ^{+0.09}_{-0.08}$ & $0.37 ^{+0.10}_{-0.07}$  \\
3 & $0.77$ & 15.58/18 & 0.71 & $0.48 ^{+0.08}_{-0.07}$  & $0.21^{+0.04}_{-0.04}$  \\
\hline
\g{4} & \g{$0.89$} & \g{20.10/18} & \g{0.27}  & \g{$0.59 ^{+0.08}_{-0.09}$} & \g{$0.19 ^{+0.07}_{-0.04}$}  \\
\g{5} & \g{$0.97$} & \g{19.68/18} & \g{0.35}  & \g{$0.64 ^{+0.08}_{-0.09}$}  & \g{$0.20 ^{+0.07}_{-0.04}$}\\
\end{tabular}
\caption{Summary table with the average redshift, best-fit chi-square to our fiducial model and number of degrees of freedom, probability to exceed (PTE), the inferred hydrostatic bias and average bias-weighted electron pressure for the different redshift bins of the \maglim sample, when cross-correlated with SPT-SZ + {\it Planck} and MILCA $y$-maps. The last two bins are shown in grey as they are not part of our fiducial sample.}
    \label{tab:gy_SNR_bPe}
\end{table}

\begin{figure*}
    \centering
    \includegraphics[width=0.45\textwidth]{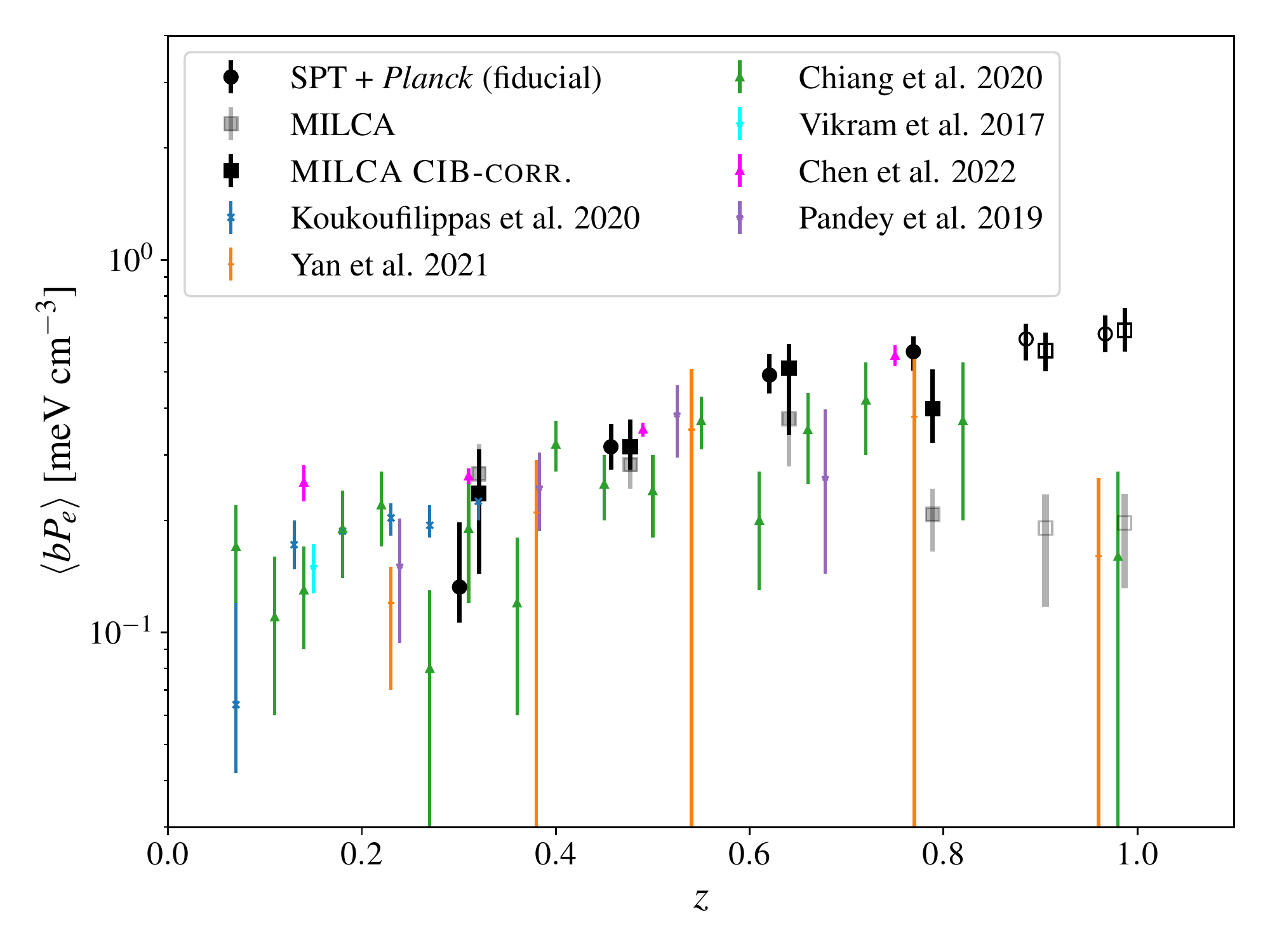}
    \includegraphics[width=0.45\textwidth]{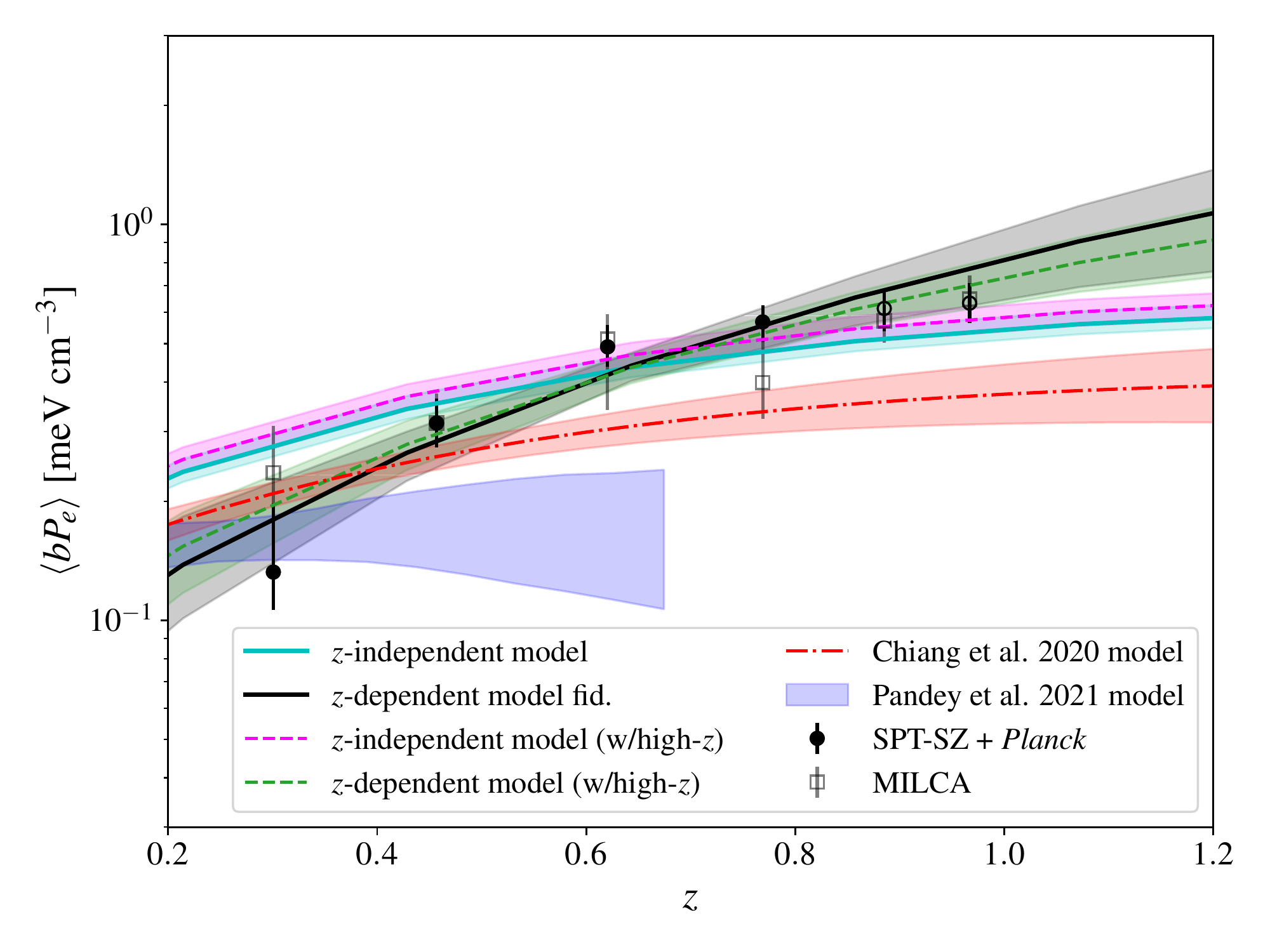}
    \includegraphics[width=0.45\textwidth]{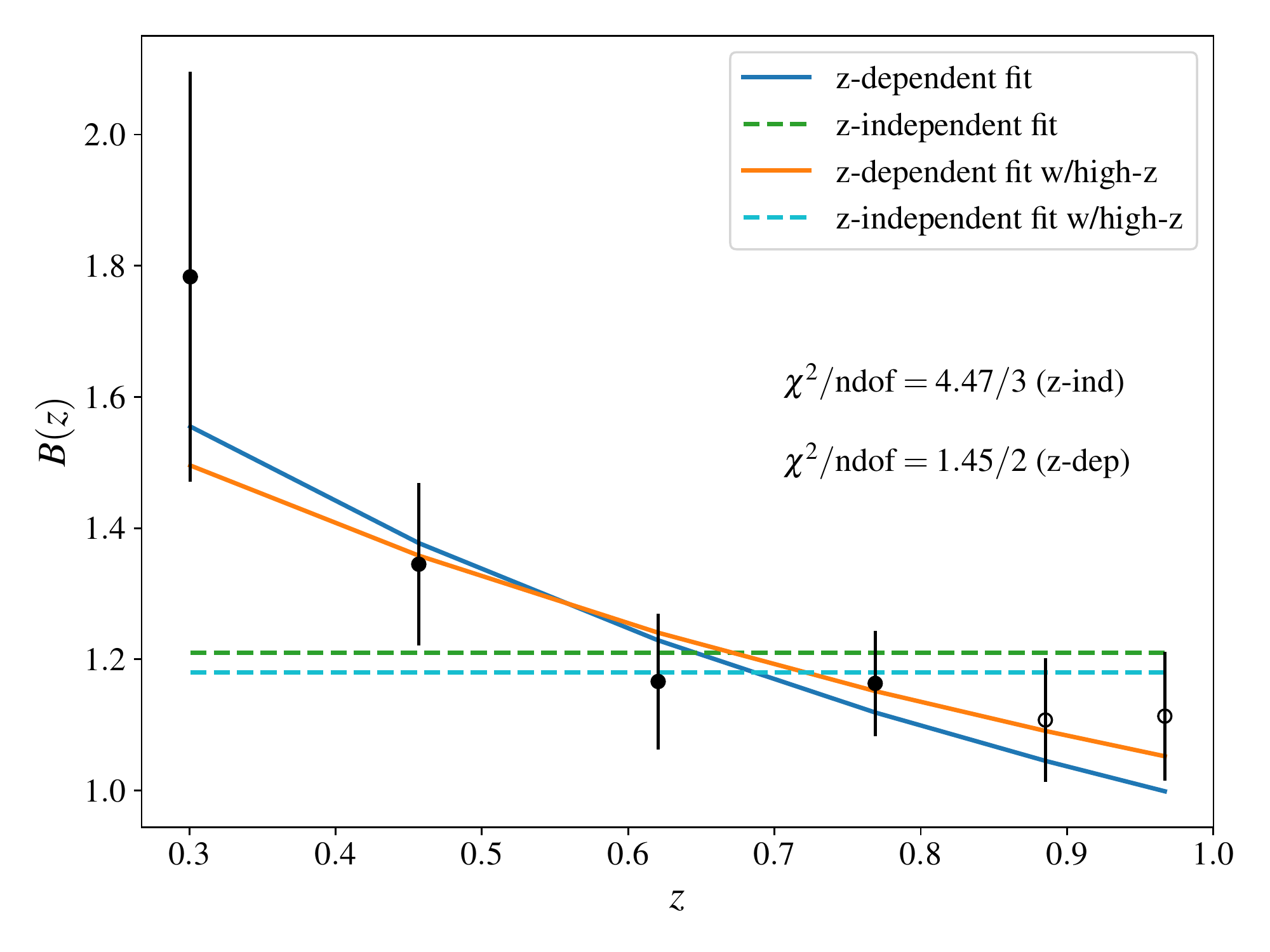}
    \caption{Top left panel: Hydrostatic bias results as a function of redshift and comparison with previous results. Top right panel: Best-fit redshift-independent model (cyan solid line), and redshift dependent model (black solid line) for the hydrostatic bias. We also show the equivalent best-fit results when including the last two redshift bins in the fits (broken magenta and green lines). For easy comparison the best-fit model from~\citet{Chiang2020} (red dot-dashed line), and from~\citet{Pandey2021} are included (shaded blue region). The model for~\citet{Pandey2021} is only included up to $z \sim 0.7$, as for higher redshifts, due to the lensing kernel, the shear $\times y$ constraining power gets reduced (Pandey private comm.). Bottom panel: Hydrostatic bias expressed as $B$ as a function of redshift, and best-fit redshift-dependent model (solid lines), and redshift-independent model (broken lines). We also include the results for the two high-redshift bins, and the best-fits including these points.}
    \label{fig:bPe_results}
\end{figure*}

The inferred best-fit values of $b_{H}$ and average bias-weighted electron pressure $\langle b_{h}P_{e} \rangle$ are summarized in Table~\ref{tab:gy_SNR_bPe}. We plot $\langle b_{h}P_{e} \rangle$ as a function of redshift in the left panel of Figure~\ref{fig:bPe_results} to compare with a compilation of previous literature \citep{Vikram2017, Pandey2019, Chiang2020, Koukoufilippas2020, Yan2021, Chen2022}. The 8 solid black points represent our fiducial result, while the 4 hollowed points are from the highest two \maglim redshift bins. Among the previous results plotted in Figure~\ref{fig:bPe_results}, \citet{Vikram2017, Pandey2019} are the most similar to this work where a combination of galaxy clustering and galaxy-$y$ cross-correlation are used and the galaxies are of similar halo mass. Overall, our results (both the fiducial sample and the high-redshift bins) appear broadly consistent with previous studies in the literature, and represent the most precise measurements at high redshift ($z>0.8$) due to the depth of  DES Y3 data.


Next we examine the redshift dependence of $\langle b_{h}P_{e} \rangle$. Following \citet{Chiang2020}, we fit $B=\frac{1}{1-b_{H}}$ as a function of redshift with a power-law model and convert the resulting fit to the corresponding model fit in $\langle b_{h}P_{e} \rangle$. In particular, we assume the following model
\begin{equation}
    B(z) = p_{0} (1+z)^{p_{1}},
\label{eq:Bofz}
\end{equation}
and fit for $p_0$ and $p_1$. Setting $p_1=0$ gives the redshift-independent model, which we also examine.

In order to account for the correlation between the data points in different redshift bins, 
we model the correlation matrix to be proportional to the overlaps in the redshift distributions between redshift bins. This is a reasonable approximation given the uncertainty in $\langle b_h P_{e} \rangle$ and therefore $B$ is dominated by the large-scale uncertainties in $C^{\rm gg}$, whose correlation between redshift bins is directly related to the overlap in the redshift bins. The 
covariance matrix then becomes:
\begin{equation}
\textrm{Cov}(B_{i}, B_{j}) = r_{ij}\sigma_{B_{i}}\sigma_{B_{j}},
\end{equation}
where $r_{ij}$ is the migration matrix given by~\citep{Benjamin2010}:
\begin{equation}
    r_{ij} = \frac{\int_{z_{i, {\rm low}}}^{z_{i, {\rm hi}}} n_{j}(z)dz}{\int_{z_{i, {\rm low}}}^{z_{i, {\rm hi}}}n_{i}(z)dz},
\label{eq:rij}
\end{equation}
where $n_{i/j}(z)$ is the normalized redshift distribution of the redshift $i/j$, and $z_{i/j, {\rm low}}, z_{i/j, {\rm hi}}$ are the photo-$z$ edges of each bin. This matrix is guaranteed to be invertible by the Gershgorin theorem as long as it is strictly diagonally dominant~\citep{Benjamin2010}, i.e.,
\begin{equation}
r_{ii} > \sum_{j \neq i} r_{ij}.
\end{equation}
Essentially, this means that the photo-$z$ distributions for each bin should be dominated by galaxies with their true redshifts within that bin. This is the case when we use the photo-z distributions in Figure~\ref{fig:zdists}.


With this procedure, we fit our data using the model of Equation~\ref{eq:Bofz}. We find that assuming a redshift independent $B$ gives $B=1.22 \pm 0.06$ with $\chi^{2}/ {\rm ndof} = 4.47 / 3$, and Bayesian evidence $\log{\mathcal{Z}}=-1.34 \pm 0.05$. On the other hand, assuming the same redshift dependent $B$ model as~\citet{Chiang2020}, we obtain $B = 2.10^{+0.61}_{-0.49}\left(1+z\right)^{-1.09^{+0.53}_{-0.53}}$ with $\chi^{2}/{\rm ndof}=1.45 / 2$, $\log{\mathcal{Z}}=-1.32 \pm 0.04$. That is, we find a Bayes' factor of 0.88, indicating no significant preference between these models. We plot both models on the upper right panel of Figure~\ref{fig:bPe_results} and compare to model fits in previous work of \citet{Chiang2020} and \citet{Pandey2021} in the bottom panel. We also show the fits including the two high-redshift bins. In general, we find that our overall constraining power on the model is at a similar level as \citet{Chiang2020}, but we prefer a decreasing trend in the values of $B$ as a function of redshift, instead of an increasing trend. This might be related to our larger values of $\langle b_{h}P_{e} \rangle$ at $z > 0.7$ compared to those found by \citet{Chiang2020} as seen in the upper left panel of Figure~\ref{fig:bPe_results}. Our findings for $p_{1}$ are in line with the results in~\citet{Wicker2022}, which find $p_{1} = -1.14 ^{+0.33}_{-0.73}$ for low mass, low redshift clusters $(z < 0.2, M < 5.89 \times 10^{14} M_{\odot})$. This decreasing trend of $B$ as a function of redshift has also been found in the literature in cluster studies~\citep{vonderLinden2014, Hoekstra2015, Smith2016, Sereno2017, Eckert2019}. We also find that when including the two highest-redshift bins, the fit does not change significantly, suggesting that even for the slightly poorer model fit in these bins and potential systematic contamination, the results using them are actually consistent with only using the fiducial sample. The comparison with \citet{Pandey2021} shows good agreement at low redshift, but there is an apparent tension at $z > 0.6$. The authors in \citet{Pandey2021} point to a lack of sensitivity to low-mass halos that contribute to the overall signal as a possible source for the tension between their work and other constraints from galaxy-$y$ cross-correlation. Moreover, due to the lensing kernel the shear-$y$ cross-correlations, the constraining power gets reduced at higher redshifts (Pandey private comm.). Future shear-$y$ cross-correlation studies using SPT data will help clarify the origin of this tension (Omori et al., in prep.). 

A by-product of this analysis that is interesting on its own is the HOD constraints of the galaxy sample, which are largely constrained by $C_{\ell}^{\rm gg}$. As only the large, 2-halo terms are used, the constraints on the 1-halo HOD parameters are weak. We nevertheless show them in Appendix~\ref{app:HOD_constraints} and compare with previous work.


\section{Small-scale analysis: constraints on gas profiles}
\label{sec:results2}


We now turn our attention to the small-scale information in the galaxy-$y$ cross-correlation. As discussed in Section~\ref{ssec:scale_cuts}, this takes advantage of the high-resolution and low-noise nature of the SPT-SZ + {\it Planck} $y$-map. We also discussed in Section~\ref{ssec:scale_cuts} that there are uncertainties in both the modeling and measurements of $C_{\ell}^{\rm gg}$. Thus, to extract the small-scale information in the gas profiles in $C_{\ell}^{\rm gy}$, we make a couple of assumptions and adjustments to the analysis. In essence, we fix the HOD constraints using the large scales as in Section~\ref{sec:results1} and Appendix~\ref{app:HOD_constraints}. Then, we fit the $C_{\ell}^{\rm gy}$ with $k_{\rm max} = 2.5$ Mpc$^{-1}$, freeing the GNFW gas profile parameters $(\alpha, \beta, \gamma)$. 

In order to get some physical intuition about our small-scale results, we compare our $C_{\ell}^{gy}$ measurements up to $k_{\rm max} = 2.5$ Mpc$^{-1}$ with the GNFW model parameters for the cosmo-OWLS suite~\citep{LeBrun14} reported in~\citet{LeBrun2015} for the ``REF'', ``AGN 8.0''\footnote{The numbers represent the AGN heating temperatures in the form of $\log_{10}({T_{\rm AGN}/K)}$ in this naming convention.}, and ``AGN 8.5'' simulations. Cosmo-OWLS is a set of hydrodynamical simulations and is part of the OverWhelmingly Large Simulations \citep[OWLS,][]{Schaye2010} project. It has been designed to help improve our understanding
of cluster astrophysics and non-linear structure formation by systematically varying several sub-grid physics models, including feedback from supernovae and AGN. One important note is that the best-fitting model in~\citet{LeBrun2015} modifies the profiles described by Equation~\ref{eq:gas_prof} by including a mass dependent concentration,  $c_{P}=c_{P,0}\left(\frac{M}{10^{14} M_{\odot}}\right)^{\delta}$, and normalization, $P_{0}=P_{0,0}\left(\frac{M}{10^{14}M_{\odot}}\right)^{\epsilon}$. The values for the parameters that we use to compare with these simulations are shown in Table~\ref{tab:sim_parameters}.

\begin{table*}
    \centering
    \begin{tabular}{cccccccc}
         Simulation & $P_{0,0}$ & $\alpha$ & $\beta$ & $\gamma$ & $c_{P,0}$ & $\delta$ & $\epsilon$ \\
         \hline
         REF & 0.528 & 2.208 & 3.632 & 1.486 & 1.192 & 0.051 & 0.210\\
         AGN 8.0 & 0.581 & 2.017 & 3.835 & 1.076 & 1.035 & 0.273 & 0.819\\
         AGN 8.5 & 0.214 & 1.868 & 4.117 & 1.063 & 0.682 & 0.245 & 0.839\\
         \hline
    \end{tabular}
    \caption{Best-fit GNFW parameters from~\citet{LeBrun2015} for different simulations of the cosmo-OWLS suite.}
    \label{tab:sim_parameters}
\end{table*}

\begin{figure*}
    \centering
    \includegraphics[width=0.95\textwidth]{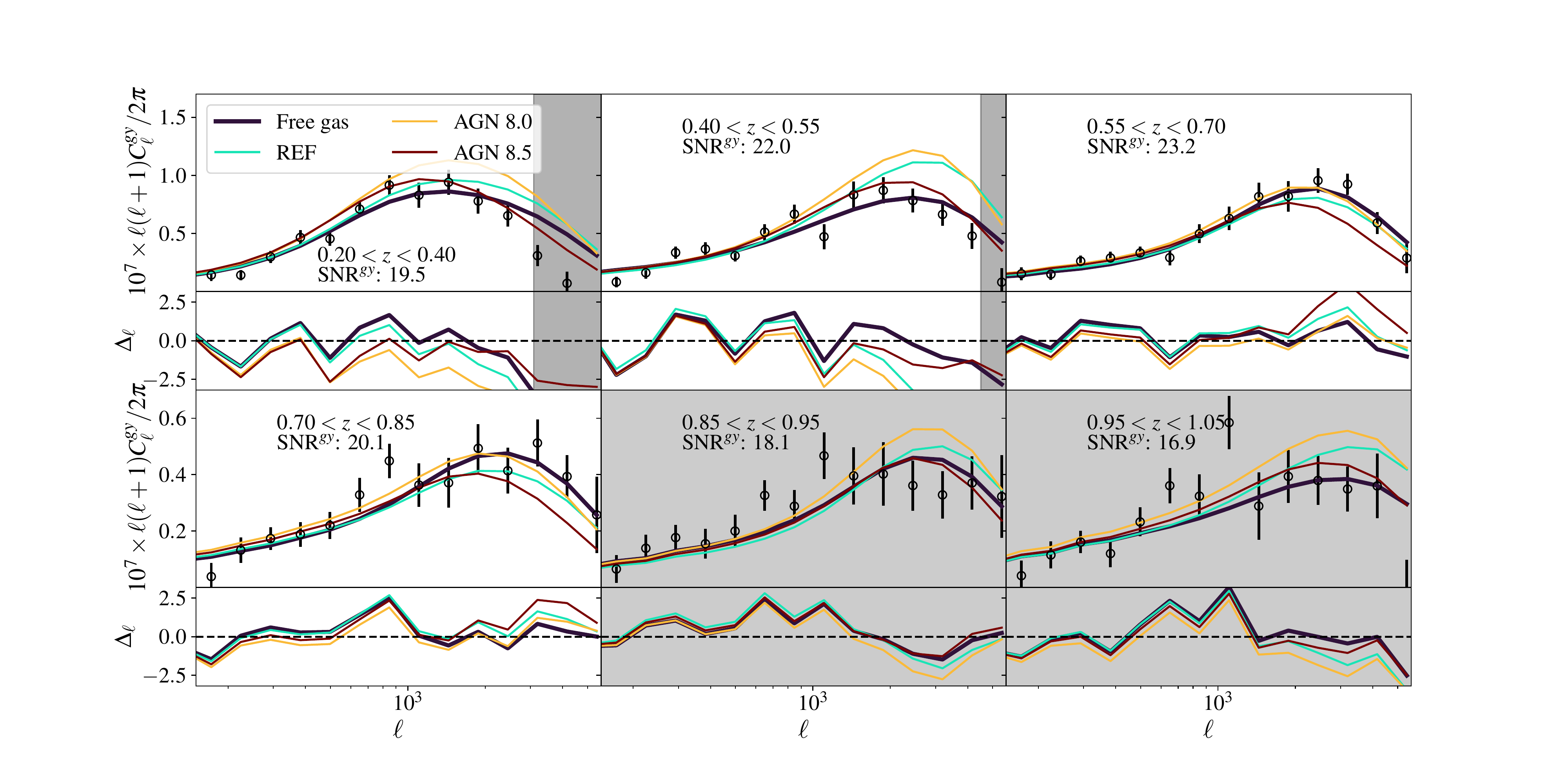}
    \caption{Top panels: Measured power spectra (black circles) and best-fit GNFW profile (black). We also include the GNFW profiles from \citet{LeBrun2015} for the REF (cyan), AGN 8.0 (orange) and AGN 8.5 (maroon) simulations of the cosmo-OWLS suite. Bottom panels: Residuals relative to the uncertainty, $\Delta_{\ell}$ for our best-fit GNFW profile (black line), for the REF model (cyan), for the AGN 8.0 model (orange), and for the AGN 8.5 model (maroon).}
    \label{fig:small_scales}
\end{figure*}

The best-fit models are shown in Figure~\ref{fig:small_scales} in black and listed in Table~\ref{tab:small_scales_results}. 
In Figure~\ref{fig:small_scales} we also overlay the prediction of the small-scale $C_{\ell}^{\rm gy}$ from different hydrodynamical simulations as listed in Table~\ref{tab:sim_parameters}. Overall, we find that for the individual redshift bins, and for the scales considered, our model provides a good description of the data and the predictions from simulations broadly follow the same trends and shape of the data vectors  

In order to quantify the overall agreement between the different models and our measurements, we combine the $\chi^2$ in the different redshift bins to a total goodness-of-fit metric
\begin{equation}
\chi^{2}_{\rm total} = \sum_{i, j=0}^{3} \sqrt{\chi_{i}^{2}} r_{ij}^{-1} \sqrt{\chi_{j}^{2}},
\label{eq:chi_tot}
\end{equation}
where $r_{ij}$ is the migration matrix defined in Equation~\ref{eq:rij}. Note that this is equivalent to considering that the only correlations between bins are due to the photometric redshift overlap, and not due to the LSS. Given our scale cuts ($\ell > 150)$ and the redshift bin width, this is a safe assumption.


We find that the GNFW model with the free gas parameters is the best description of our data with $\chi^{2}_{\rm total}/\rm{ndof} = 42/40$ (PTE: 0.37), followed by the AGN 8.5 model $\chi^{2}_{\rm total} /\rm{ndof} = 58/52$ (PTE: 0.25).\footnote{We find that for some of the bins, the model using the parameters from the AGN 8.5 simulation have a lower $\chi^{2}$ than our model. This is due to added flexibility in the model by adding a mass dependency in $P_{0}$ and $c_{P}$.} The REF simulation is a reasonable but somewhat worse fit with $\chi^{2}_{\rm total}/\rm{ndof} = 84/52$ (PTE: 0.003) -- we note that we generally expect REF to perform worse than the the AGN models as it does not contain \textit{any} AGN feedback, which is not realistic, as we expect AGN feedback to be very relevant for the observables considered here. Finally, the AGN 8.0 model seems to poorly describe bins 0, and 1, where the latter has a $\chi^{2}/\rm{ndof} = 76.8/13$. The total $\chi^{2}_{\rm total}/\rm{ndof} = 108/52$ (PTE:$6 \times 10^{-6}$) indicates that this model does not describe the data well. 

Taken at face value, the data prefers the AGN 8.5 model over the AGN 8.0 model, indicating that stronger feedback (higher $T_{AGN}$) is preferred. This is opposite from what is found in \citep{Troester21KIDS} using cosmic shear only.
However, our results do 
agree with 
several other 
previous work including \citet{Gatti2021}, \citet{Lim2018} and \citet{Troester21}. It is worth noting that there are some results in the literature that use the BAHAMAS~\citep{McCarthy2017} simulations model, and the value for best-fit AGN heating temperature found using this model is different than using the cosmo-OWLS model as is done here. For example, the authors in~\citet{McCarthy2017} point out that cosmo-OWLS AGN  8.0 matches best with AGN 7.8 for BAHAMAS.

\begin{figure}
    \centering
    \includegraphics[width=0.95\columnwidth]{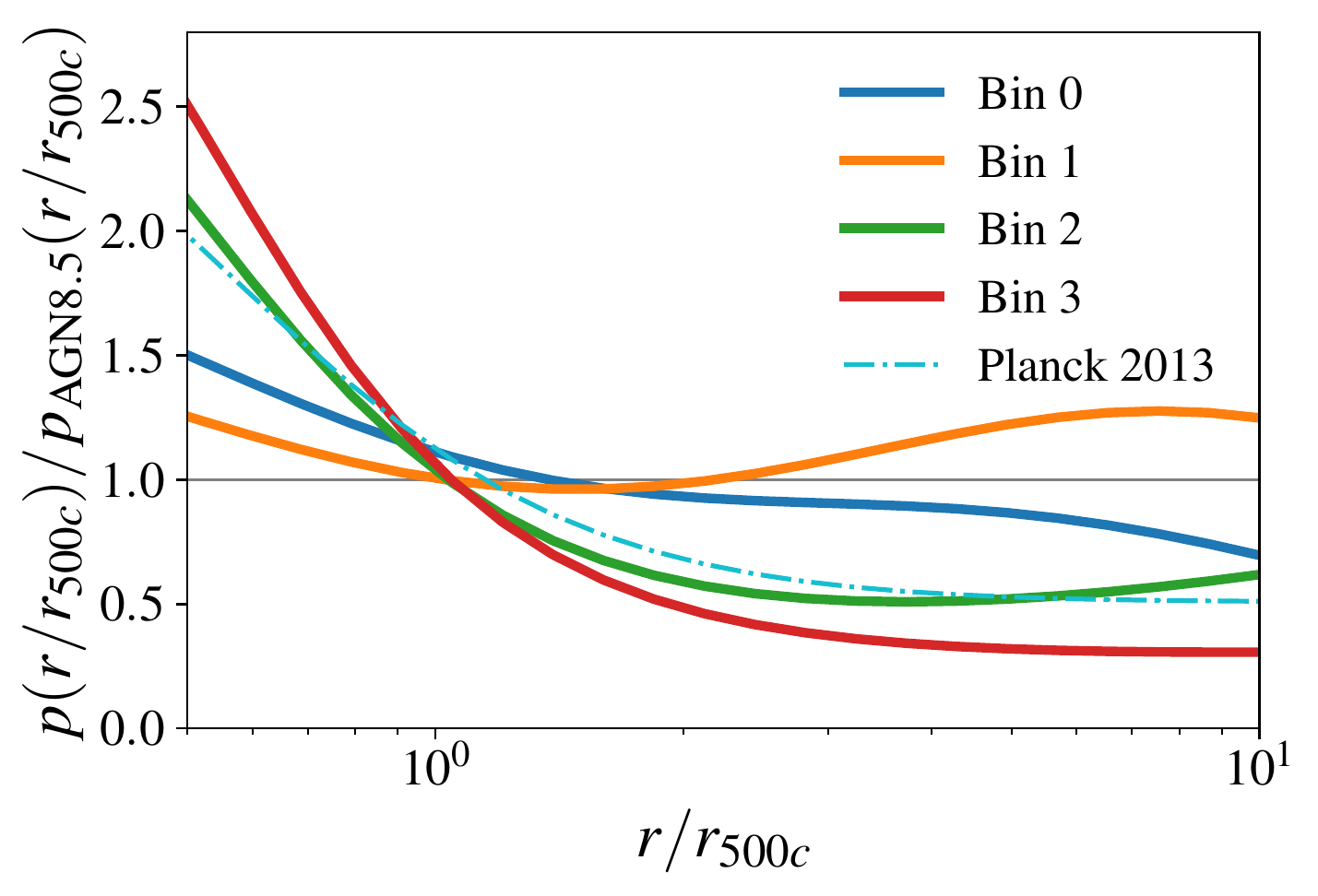}
    \caption{Ratio of the best-fit gas profiles from the four tomographic bins in our work (solid lines) to the ``AGN 8.5'' simulation in~\citet{LeBrun2015}. We also plot best-fit profile to Planck clusters (dot dashed cyan).}
    \label{fig:profiles}
\end{figure}

Our results also seem to indicate that, at low redshifts, the AGN feedback is stronger than at higher redshifts (both AGN 8.0 and AGN 8.5 lines are systematically above the data points in bin 0 and 1, and similar or below the data points in bin 2 and 3), pointing to potential redshift-dependent AGN feedback amplitudes that are somewhat different compared to the cosmo-OWLS simulations. Furthermore, for the first two redshift bins, our GNFW fits prefer smaller values for $\alpha, c_{P}$ and larger values for $\beta$ than those found for galaxy clusters in~\citet{2010A&A...517A..92A} and~\citet{Planck2013a}, which means that the gas profiles for our galaxy sample are more peaked than typical cluster profiles. We show the ratio of the best-fit pressure profile to the AGN 8.5 profile at $M=10^{14}M_{\odot}/h$ in Figure~\ref{fig:profiles}. It is clear from this plot that, in general, the data prefer a profile with higher pressure at small scales $r< r_{500c}$ compared to the AGN 8.5 profile. In addition, there is a clear redshift evolution -- this difference is most apparent at high redshift. We also include the best-fit profile from {\it Planck} galaxy clusters, which shows a similar trend as our results and interestingly matches well our results from Bin 2.



We note that given the small number of simulations we are able to compare to and the uncertainties in the data, we cannot make more generic astrophysical constraints on, e.g., the AGN heating temperature. We can, however, conclude that in the cosmo-OWLS simulation suite, the AGN 8.5 case, with all parameter settings adopted by that simulation, can describe our data within the measurement uncertainties. This will suggest that in future analyses when a plausible small-scale model is needed for, e.g., the matter power spectrum, AGN 8.5 will be a good option. We leave it for future work to apply the methodology of this work to more recent simulation suits that systematically span a larger grid of different baryonic physics such as the Cosmology and Astrophysics with Machine-learning Simulations project \citep[CAMELS,][]{VillaescusaNavarro2021}. 


\begin{table*}
    \centering
    \begin{tabular}{ccccccccc}
        Bin &  $n_{\rm data}$ & $\alpha$ & $\beta$ & $c_{P}$ & $\chi^{2}$ (free gas) & $\chi^{2}$ (REF) &
        $\chi^{2}$ (AGN 8.0) & $\chi^{2}$ (AGN 8.5)\\
        \hline
        0 & 11 & $0.74^{+0.33}_{-0.13}$ & $5.67^{+2.23}_{-1.81}$ & $0.58^{+1.17}_{-0.38}$ 
        & 9.7 & 13.6 & 26.6 & 10.5\\
        1 & 13 & $0.62^{+0.14}_{-0.08}$ &
        $5.89^{+1.61}_{-1.38}$ & $0.38^{+0.55}_{-0.22}$ & 22.5 & 62.0 & 76.8 & 23.2\\
        2 & 14  & $1.78^{+1.30}_{-0.70}$ &
        $3.73^{+1.25}_{-0.43}$ & $2.38^{+0.67}_{-1.18}$ & 6.1 & 6.7 & 6.3 & 15.9\\
        3 & 14  & $2.01^{+1.25}_{-0.81}$ & $4.03^{+1.26}_{-0.53}$ & $2.24^{+0.60}_{-0.97}$
        & 11.3 & 14.1 & 14.8 & 19.8\\
        \hline
        \g{4} & \g{14} & \g{$1.80^{+1.44}_{-0.92}$} & 
        \g{$3.06^{+0.75}_{-0.28}$} & \g{$3.36^{+0.78}_{-1.75}$} & \g{19.2} & \g{29.4} & \g{25.8} & \g{23.0}\\
        \g{5} & \g{14} & \g{$1.38^{+1.50}_{-0.62}$} 
        & \g{$3.16^{+0.94}_{-0.34}$} & \g{$3.16^{+1.10}_{-1.87}$} & \g{21.5} & \g{26.4} & \g{34.3} & \g{20.1}\\
        \hline
    \end{tabular}
    \caption{Summary of the number of data points, the best-fit parameters and the $\chi^2$ values against predictions from different simulations for the small-scale measurements in the SPT region described in Section~\ref{sec:results2}. The last two bins are shown in grey as they are not part of our fiducial sample. }
    \label{tab:small_scales_results}
\end{table*}

\section{Conclusions}
\label{sec:conclusions}

In this paper we measure and model the auto and cross-power spectra of galaxy density and the thermal Sunyaev-Zeldovich (tSZ) effect. We use the \maglim galaxies from the first 3 years of DES data and the MILCA $y$-map, as well the $y$-map constructed using frequency combinations from SPT-SZ and {\it Planck}. There are two main goals of this paper: First we use the large-scale information to constrain the hydrostatic mass bias and compare with previous studies. Second, fixing the HOD from the large-scale galaxy density auto- correlation, we fit the galaxy density-$y$ cross-correlation and extract information about the small-scale gas profile. We then compare our results with hydrodynamical simulations.

The main advances of this work are the high density of the \maglim sample, which allows us to have a good SNR from redshift 0.2 to 1.0, and the new $y$-maps from SPT, which probe smaller scales than previous works. Additionally, having three different versions of the SPT $y$-maps (minimum variance, CMB-CIB-nulled and CIB-nulled) allows us a better understanding of the main foreground at $z > 0.7$, the contribution from the Cosmic Infrared Background (CIB).

We model the power spectra using an HOD approach, finding good agreement with previous results~\citep{Zacharegkas2021}. We find that our measurements of the hydrostatic bias, $b_{H}$, are in good agreement with the literature~\citep{Vikram2017, Pandey2019, Chiang2020, Koukoufilippas2020, Yan2021, Chen2022}. 

We measure the evolution of $b_{H}$ with redshift, and find a slight preference for a redshift dependent $b_{H}$, and find the evolution to be somewhat stronger than previous results. Additional measurements at higher and lower redshifts should be used to pin down the evolution in future work. 
Our results using the \textit{Planck} MILCA maps for $z \gtrsim 0.8$ show CIB contamination levels of 70\% 
in the measured $b_{H}$. As a result, these maps should be used with caution for cross-correlation studies. 
We find that the bias due to the presence of CIB  in the MILCA $y$-map can be mitigated by estimating the contamination term (by propagating external estimated of the CIB at various frequencies through the MILCA weights), and subtracting that contribution from the measured correlation.

Finally, we take advantage of the high-resolution nature of the SPT maps and fit the measured gas profiles to a GNFW profile, and compare with standard profiles from the cosmo-OWLS suite~\citep{LeBrun2015}. We find that, overall, our results seem to prefer the strongest of the considered AGN feedback models, AGN 8.5. This agrees with previous results using shear-$y$ cross-correlations~\citep{Gatti2021, Troester21}. We also notice that this preference for higher-than-typical AGN feedback is lowered when going to higher redshifts, which may indicate a redshift evolution. 

Upcoming data from Advanced ACTPol~\citep{AdvACT} and SPT-3G~\citep{SPT-3G}, in combination with the final DES dataset and the Hyper Suprime-Cam Subaru Strategic Program~\citep[HSC-SSP][]{HSC-survey} can help us further study this potential evolution, and improve the modeling of the small scales, which constitute one of the main limiting factors of current large-scale structure cosmology analyses.

\section*{Acknowledgements}
JS thanks Yi-Kuan Chiang for providing $\langle b_{h}P_{e} \rangle$ data for comparisons. JS thanks David Alonso for help and feedback throughout the project. 
We thank the creators of \texttt{CCL, astropy, numpy, matplotlib, yxg}.

YO and CC are supported by DOE grant DE-SC0021949.

The South Pole Telescope program is supported by the
National Science Foundation (NSF) through the grant
OPP-1852617 and OPP-2147371. Partial support is also provided by the
Kavli Institute of Cosmological Physics at the University of Chicago. Argonne National Laboratory's work
was supported by the U.S. Department of Energy, Office
of Science, Office of High Energy Physics, under contract DE-AC02-06CH11357. Work at Fermi National
Accelerator Laboratory, a DOE-OS, HEP User Facility managed by the Fermi Research Alliance, LLC, was
supported under Contract No. DE-AC02- 07CH11359.
The Melbourne authors acknowledge support from the
Australian Research Council's Discovery Projects scheme
(DP200101068). The McGill authors acknowledge funding from the Natural Sciences and Engineering Research Council of Canada, Canadian Institute for Advanced research, and the Fonds de recherche du Qu\'{e}bec
Nature et technologies. The CU Boulder group acknowledges support from NSF AST-0956135. The Munich group acknowledges the support by the ORIGINS
Cluster (funded by the Deutsche Forschungsgemeinschaft (DFG, German Research Foundation) under Germany’s Excellence Strategy – EXC-2094 – 390783311),
the MaxPlanck-Gesellschaft Faculty Fellowship Program,
and the Ludwig-Maximilians-Universität M\"{u}nchen.

Funding for the DES Projects has been provided by the U.S. Department of Energy, the U.S. National Science Foundation, the Ministry of Science and Education of Spain, 
the Science and Technology Facilities Council of the United Kingdom, the Higher Education Funding Council for England, the National Center for Supercomputing 
Applications at the University of Illinois at Urbana-Champaign, the Kavli Institute of Cosmological Physics at the University of Chicago, 
the Center for Cosmology and Astro-Particle Physics at the Ohio State University,
the Mitchell Institute for Fundamental Physics and Astronomy at Texas A\&M University, Financiadora de Estudos e Projetos, 
Funda{\c c}{\~a}o Carlos Chagas Filho de Amparo {\`a} Pesquisa do Estado do Rio de Janeiro, Conselho Nacional de Desenvolvimento Cient{\'i}fico e Tecnol{\'o}gico and 
the Minist{\'e}rio da Ci{\^e}ncia, Tecnologia e Inova{\c c}{\~a}o, the Deutsche Forschungsgemeinschaft and the Collaborating Institutions in the Dark Energy Survey. 

The Collaborating Institutions are Argonne National Laboratory, the University of California at Santa Cruz, the University of Cambridge, Centro de Investigaciones Energ{\'e}ticas, 
Medioambientales y Tecnol{\'o}gicas-Madrid, the University of Chicago, University College London, the DES-Brazil Consortium, the University of Edinburgh, 
the Eidgen{\"o}ssische Technische Hochschule (ETH) Z{\"u}rich, 
Fermi National Accelerator Laboratory, the University of Illinois at Urbana-Champaign, the Institut de Ci{\`e}ncies de l'Espai (IEEC/CSIC), 
the Institut de F{\'i}sica d'Altes Energies, Lawrence Berkeley National Laboratory, the Ludwig-Maximilians Universit{\"a}t M{\"u}nchen and the associated Excellence Cluster Universe, 
the University of Michigan, NFS's NOIRLab, the University of Nottingham, The Ohio State University, the University of Pennsylvania, the University of Portsmouth, 
SLAC National Accelerator Laboratory, Stanford University, the University of Sussex, Texas A\&M University, and the OzDES Membership Consortium.

Based in part on observations at Cerro Tololo Inter-American Observatory at NSF's NOIRLab (NOIRLab Prop. ID 2012B-0001; PI: J. Frieman), which is managed by the Association of Universities for Research in Astronomy (AURA) under a cooperative agreement with the National Science Foundation.

The DES data management system is supported by the National Science Foundation under Grant Numbers AST-1138766 and AST-1536171.
The DES participants from Spanish institutions are partially supported by MICINN under grants ESP2017-89838, PGC2018-094773, PGC2018-102021, SEV-2016-0588, SEV-2016-0597, and MDM-2015-0509, some of which include ERDF funds from the European Union. IFAE is partially funded by the CERCA program of the Generalitat de Catalunya.
Research leading to these results has received funding from the European Research
Council under the European Union's Seventh Framework Program (FP7/2007-2013) including ERC grant agreements 240672, 291329, and 306478.
We  acknowledge support from the Brazilian Instituto Nacional de Ci\^encia
e Tecnologia (INCT) do e-Universo (CNPq grant 465376/2014-2).

This manuscript has been authored by Fermi Research Alliance, LLC under Contract No. DE-AC02-07CH11359 with the U.S. Department of Energy, Office of Science, Office of High Energy Physics.

We gratefully acknowledge the computing resources provided on Crossover, a high-performance computing cluster operated by the Laboratory Computing Resource Center at Argonne National Laboratory.

This research used resources of the National Energy Research Scientific Computing Center (NERSC), a U.S. Department of Energy Office of Science User Facility located at Lawrence Berkeley National Laboratory, operated under Contract No. DE-AC02-05CH11231.

\section*{Affiliations}
$^{1}$ Space Telescope Science Institute, 3700 San Martin Drive, Baltimore, MD 21210, USA\\
$^{2}$ Fermi National Accelerator Laboratory, P. O. Box 500, Batavia, IL 60510, USA\\
$^{3}$ Kavli Institute for Cosmological Physics, University of Chicago, Chicago, IL 60637, USA\\
$^{4}$ Department of Astronomy and Astrophysics, University of Chicago, Chicago, IL 60637, USA\\
$^{5}$ Department of Physics, Stanford University, 382 Via Pueblo Mall, Stanford, CA 94305, USA\\
$^{6}$ Kavli Institute for Particle Astrophysics \& Cosmology, P. O. Box 2450, Stanford University, Stanford, CA 94305, USA\\
$^{7}$ High-Energy Physics Division, Argonne National Laboratory, 9700 South Cass Avenue, Argonne, IL 60439, USA\\
$^{8}$ Center for Astrophysical Surveys, National Center for Supercomputing Applications, 1205 West Clark St., Urbana, IL 61801, USA\\
$^{9}$ Cerro Tololo Inter-American Observatory, NSF's National Optical-Infrared Astronomy Research Laboratory, Casilla 603, La Serena, Chile\\
$^{10}$ Laborat\'orio Interinstitucional de e-Astronomia - LIneA, Rua Gal. Jos\'e Cristino 77, Rio de Janeiro, RJ - 20921-400, Brazil\\
$^{11}$ Department of Physics, University of Michigan, Ann Arbor, MI 48109, USA\\
$^{12}$ Kavli Institute for Cosmology Cambridge, Madingley Road, Cambridge, CB3 OHA, UK\\
$^{13}$ Instituto de Fisica Teorica UAM/CSIC, Universidad Autonoma de Madrid, 28049 Madrid, Spain\\
$^{14}$ Institute for Astronomy, University of Hawai'i, 2680 Woodlawn Drive, Honolulu, HI 96822, USA\\
$^{15}$ Physics Department, 2320 Chamberlin Hall, University of Wisconsin-Madison, 1150 University Avenue Madison, WI  53706-1390\\
$^{16}$ Department of Physics and Astronomy, University of Pennsylvania, Philadelphia, PA 19104, USA\\
$^{17}$ CNRS, UMR 7095, Institut d'Astrophysique de Paris, F-75014, Paris, France\\
$^{18}$ Sorbonne Universit\'es, UPMC Univ Paris 06, UMR 7095, Institut d'Astrophysique de Paris, F-75014, Paris, France\\
$^{19}$ University Observatory, Faculty of Physics, Ludwig-Maximilians-Universit\"at, Scheinerstr. 1, 81679 Munich, Germany\\
$^{20}$ Department of Physics \& Astronomy, University College London, Gower Street, London, WC1E 6BT, UK\\
$^{21}$ SLAC National Accelerator Laboratory, Menlo Park, CA 94025, USA\\
$^{22}$ Department of Physics, Carnegie Mellon University, Pittsburgh, Pennsylvania 15312, USA\\
$^{23}$ Enrico Fermi Institute, University of Chicago, Chicago, IL 60637, USA\\
$^{24}$ Department of Physics, University of Chicago, Chicago, IL 60637, USA\\
$^{25}$ Instituto de Astrof\'{i}sica de Canarias, E-38205 La Laguna, Tenerife, Spain\\
$^{26}$ Universidad de La Laguna, Dpto. Astrof\'{i}sica, E-38206 La Laguna, Tenerife, Spain\\
$^{27}$ Department of Astronomy, University of Illinois Urbana-Champaign, 1002 West Green Street, Urbana, IL, 61801, USA\\
$^{28}$ Institut de F\'{\i}sica d'Altes Energies (IFAE), The Barcelona Institute of Science and Technology, Campus UAB, 08193 Bellaterra (Barcelona) Spain\\
$^{29}$ Institut d'Estudis Espacials de Catalunya (IEEC), 08034 Barcelona, Spain\\
$^{30}$ Institute of Space Sciences (ICE, CSIC),  Campus UAB, Carrer de Can Magrans, s/n,  08193 Barcelona, Spain\\
$^{31}$ Physics Department, William Jewell College, Liberty, MO, 64068\\
$^{32}$ Kavli Institute for the Physics and Mathematics of the Universe (WPI), UTIAS, The University of Tokyo, Kashiwa, Chiba 277-8583, Japan\\
$^{33}$ NASA Goddard Space Flight Center, 8800 Greenbelt Rd, Greenbelt, MD 20771, USA \\
$^{34}$ Department of Physics \& Astronomy, The University of Western Ontario, London ON N6A 3K7, Canada\\
$^{35}$ Institute for Earth and Space Exploration, The University of Western Ontario, London ON N6A 3K7, Canada\\
$^{36}$ Astronomy Unit, Department of Physics, University of Trieste, via Tiepolo 11, I-34131 Trieste, Italy\\
$^{37}$ INAF-Osservatorio Astronomico di Trieste, via G. B. Tiepolo 11, I-34143 Trieste, Italy\\
$^{38}$ Institute for Fundamental Physics of the Universe, Via Beirut 2, 34014 Trieste, Italy\\
$^{39}$ California Institute of Technology, 1200 East California Blvd, MC 249-17, Pasadena, CA 91125, USA\\
$^{40}$ Hamburger Sternwarte, Universit\"{a}t Hamburg, Gojenbergsweg 112, 21029 Hamburg, Germany\\
$^{41}$ High Energy Accelerator Research Organization (KEK), Tsukuba, Ibaraki 305-0801, Japan\\
$^{42}$ Department of Physics, University of California, Berkeley, CA, 94720, USA\\
$^{43}$ Centro de Investigaciones Energ\'eticas, Medioambientales y Tecnol\'ogicas (CIEMAT), Madrid, Spain\\
$^{44}$ Lawrence Berkeley National Laboratory, 1 Cyclotron Road, Berkeley, CA 94720, USA\\
$^{45}$ Department of Physics, IIT Hyderabad, Kandi, Telangana 502285, India\\
$^{46}$ Department of Physics and McGill Space Institute, McGill University, 3600 Rue University, Montreal, Quebec H3A 2T8, Canada\\
$^{47}$ Canadian Institute for Advanced Research, CIFAR Program in Gravity and the Extreme Universe, Toronto, ON, M5G 1Z8, Canada\\
$^{48}$ NSF AI Planning Institute for Physics of the Future, Carnegie Mellon University, Pittsburgh, PA 15213, USA\\
$^{49}$ Center for Cosmology and Astro-Particle Physics, The Ohio State University, Columbus, OH 43210, USA\\
$^{50}$ Department of Physics, The Ohio State University, Columbus, OH 43210, USA\\
$^{51}$ Department of Astrophysical and Planetary Sciences, University of Colorado, Boulder, CO, 80309, USA\\
$^{52}$ Jet Propulsion Laboratory, California Institute of Technology, 4800 Oak Grove Dr., Pasadena, CA 91109, USA\\
$^{53}$ Institute of Theoretical Astrophysics, University of Oslo. P.O. Box 1029 Blindern, NO-0315 Oslo, Norway\\
$^{54}$ European Southern Observatory, Karl-Schwarzschild-Straße 2, 85748 Garching, Germany\\
$^{55}$ Department of Astronomy, University of Michigan, Ann Arbor, MI 48109, USA\\
$^{56}$ Observat\'orio Nacional, Rua Gal. Jos\'e Cristino 77, Rio de Janeiro, RJ - 20921-400, Brazil\\
$^{57}$ Department of Physics, University of Colorado, Boulder, CO, 80309, USA\\
$^{58}$ School of Mathematics and Physics, University of Queensland,  Brisbane, QLD 4072, Australia\\
$^{59}$ Department of Physics, University of Illinois Urbana-Champaign, 1110 West Green Street, Urbana, IL 61801, USA\\
$^{60}$ Santa Cruz Institute for Particle Physics, Santa Cruz, CA 95064, USA\\
$^{61}$ University of Chicago, Chicago, IL 60637, USA\\
$^{62}$ Center for Astrophysics $\vert$ Harvard \& Smithsonian, 60 Garden Street, Cambridge, MA 02138, USA\\
$^{63}$ Department of Physics, University of California, One Shields Avenue, Davis, CA, 95616, USA\\
$^{64}$ Australian Astronomical Optics, Macquarie University, North Ryde, NSW 2113, Australia\\
$^{65}$ Lowell Observatory, 1400 Mars Hill Rd, Flagstaff, AZ 86001, USA\\
$^{66}$ Physics Division, Lawrence Berkeley National Laboratory, Berkeley, CA, 94720, USA\\
$^{67}$ Department of Applied Mathematics and Theoretical Physics, University of Cambridge, Cambridge CB3 0WA, UK\\
$^{68}$ George P. and Cynthia Woods Mitchell Institute for Fundamental Physics and Astronomy, and Department of Physics and Astronomy, Texas A\&M University, College Station, TX 77843,  USA\\
$^{69}$ Department of Astrophysical Sciences, Princeton University, Peyton Hall, Princeton, NJ 08544, USA\\
$^{70}$ Instituci\'o Catalana de Recerca i Estudis Avan\c{c}ats, E-08010 Barcelona, Spain\\
$^{71}$ Excellence Cluster Universe, Boltzmannstr.\ 2, 85748 Garching, Germany\\
$^{72}$ Ludwig-Maximilians-Universit{\"a}t, Scheiner- str. 1, 81679 Munich, Germany\\
$^{73}$ Max-Planck-Institut f\"{u}r extraterrestrische Physik,Giessenbachstr.\ 85748 Garching, Germany\\
$^{74}$ Perimeter Institute for Theoretical Physics, 31 Caroline St. North, Waterloo, ON N2L 2Y5, Canada\\
$^{75}$ Dunlap Institute for Astronomy \& Astrophysics, University of Toronto, 50 St. George Street, Toronto, ON, M5S 3H4, Canada\\
$^{76}$ Department of Astronomy, University of California, Berkeley,  501 Campbell Hall, Berkeley, CA 94720, USA\\
$^{77}$ Institute of Astronomy, University of Cambridge, Madingley Road, Cambridge CB3 0HA, UK\\
$^{78}$ Institute for Astronomy, University of Edinburgh, Edinburgh EH9 3HJ, UK\\
$^{79}$ School of Physics and Astronomy, University of Minnesota, 116 Church Street SE Minneapolis, MN, 55455, USA\\
$^{80}$ Department of Physics, University or Genova and INFN, Genova division, Via Dodecaneso 33, 16146, Genova, Italy\\
$^{81}$ School of Physics, University of Melbourne, Parkville, VIC 3010, Australia\\
$^{82}$ Department of Physics, Case Western Reserve University, Cleveland, OH, 44106, USA\\
$^{83}$ Liberal Arts Department, School of the Art Institute of Chicago, Chicago, IL USA 60603\\
$^{84}$ Brookhaven National Laboratory, Bldg 510, Upton, NY 11973, USA\\
$^{85}$ School of Physics and Astronomy, University of Southampton,  Southampton, SO17 1BJ, UK\\
$^{86}$ Computer Science and Mathematics Division, Oak Ridge National Laboratory, Oak Ridge, TN 37831\\
$^{87}$ Institute of Cosmology and Gravitation, University of Portsmouth, Portsmouth, PO1 3FX, UK\\
$^{88}$ Department of Physics, Duke University Durham, NC 27708, USA\\

\section*{Data Availability}
All DES Year 3 cosmology catalogs are available at \url{https://des.ncsa.illinois.edu/releases/y3a2}. This work uses publicly available SPT-SZ+\Planck maps at \url{https://pole.uchicago.edu/public/data/sptsz_ymap/} and the \Planck MILCA map at \url{https://irsa.ipac.caltech.edu/data/Planck/release_2/all-sky-maps/ysz_index.html}. In addition we use the publicly available CIB maps from \url{https://lambda.gsfc.nasa.gov/product/planck/curr/planck_tp_lenz_get.html}


\bibliographystyle{mnras}
\bibliography{bibliography, y3kp} 


\appendix

\section{HOD constraints from galaxy clustering}
\label{app:HOD_constraints}

\begin{figure}
    \centering
    \includegraphics[width=0.9\columnwidth]{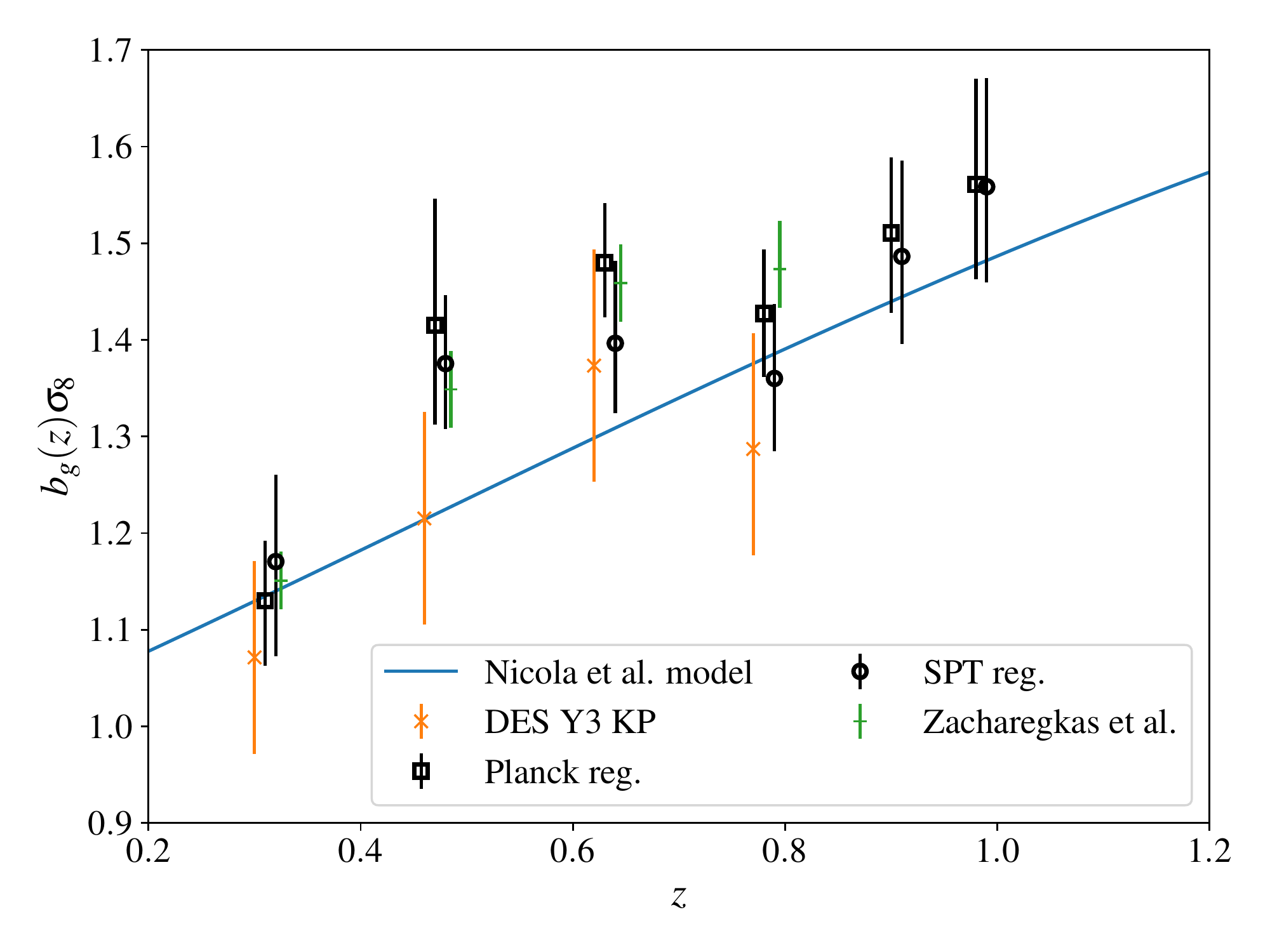}
    \caption{Galaxy bias results from the SPT (open circles), Planck (open squares) regions, compared to the DES Y3 results~\citep{DES2021}, the results from~\citep{Zacharegkas2021}, and with the model from~\citet{Nicola2020}.}
    \label{fig:galaxy_bias_evo}
\end{figure}

As a by-product of this analysis, we obtain the HOD constraints from galaxy clustering for the DES Y3 \maglim sample. A dedicated HOD analysis of the same sample using a galaxy-galaxy lensing approach was presented in \citet{Zacharegkas2021}. Being a cross-correlation, galaxy-galaxy lensing is somewhat less prone to systematic effects. As a result checking the HOD constraints from our analysis using galaxy clustering with \citet{Zacharegkas2021} provides a good test for the robustness of the constraints. In particular, a good test is to compare the galaxy bias obtained using the best-fit halo model parameters from this work, with the galaxy bias results measured in \citet{Zacharegkas2021}. We also compare these bias results with the results obtained in the $3\times 2$-point DES Y3 analysis~\citep{DES2021} (DES Y3 KP), and the model in~\citet{Nicola2020}:

\begin{equation}
    b(m_{\rm{lim}}, z) = \left[b_{1}\left(m_{\rm{lim}} - 24 \right) + b_{0}\right]D_{+}(z)^{\alpha},
\end{equation}
With $b_{1} = -0.0624 \pm 0.0070$, $b_{0} = 0.8346 \pm 0.161$, $\alpha = -1.30 \pm 0.19$, and $D_{+}(z)$ the linear growth factor at redshift $z$. In our case we use $m_{\rm{lim}} = 4 z + 18$. Since our measurements of $C_{\ell}^{gg}$ are only sensitive to the $b_{g}\sigma_{8}$ combination, we decide to compare on these terms instead of the raw galaxy bias, as some of these analyses use slightly different values for $\sigma_{8}$. In Figure~\ref{fig:galaxy_bias_evo} we see that all of the results using the DES Y3 are in good agreement. We also see that the results from the SPT and \textit{Planck} regions agree with each other. Finally, we see that our measurements are compatible with the model from~\citet{Nicola2020}, despite using a different galaxy sample (DES Y3 vs HSC). We include the best-fit HOD parameters in Table~\ref{tab:HOD_results} for reference.

\begin{table*}
    \centering
    \begin{tabular}{cccccccccc}
    \hline
        Bin & $\log_{10}M_{\rm min}/M_{\odot}$ & $\log_{10}M'_{1}/M_{\odot}$ & $\alpha_{s}$ & $\beta_{g}$ & $\beta_{\rm max}$ & $\rho_{gy}$ & $\Delta z_{i}$ & $\sigma_{z,i}$\\
        \hline
        \multicolumn{9}{c}{SPT} \\
        \hline
         0 & $12.1^{+0.3}_{-0.4}$ & $13.2^{+0.5}_{-1.0}$ & $0.97^{+0.27}_{-0.22}$ & $1.0^{+0.2}_{-0.5}$ & $2.3^{+2.0}_{-0.8}$ & $0.21^{+0.50}_{-0.40}$ & $\left(-9.0 ^{+7.7}_{-7.1}\right)\times 10^{-3}$ & $0.97^{+0.06}_{-0.06}$\\
         1 & $12.1^{+0.3}_{-0.3}$ & $13.1^{+0.8}_{-1.3}$ & $0.62^{+0.19}_{-0.29}$ & $1.3^{+0.2}_{-0.1}$ & $0.9^{+1.1}_{-0.5}$ & $-0.28^{+0.65}_{-0.25}$ & $\left(-3.4 ^{+1.2}_{-1.1}\right)\times 10^{-2}$ & $1.31^{+0.09}_{-0.10}$\\
         2 & $12.1^{+0.3}_{-0.3}$ & $12.9^{+0.6}_{-1.2}$ & $0.71^{+0.32}_{-0.35}$ & $1.2^{+0.1}_{-0.1}$ & $1.9^{+2.6}_{-1.1}$ & $-0.30^{+0.65}_{-0.28}$ & $\left(-4.3 ^{+5.9}_{-6.6}\right)\times 10^{-3}$ & $0.88^{+0.05}_{-0.05}$\\
         3 & $12.1^{+0.2}_{-0.2}$ & $12.7^{+0.6}_{-1.1}$ & $0.54^{+0.52}_{-0.31}$ & $1.1^{+0.2}_{-0.3}$ & $4.0^{+3.6}_{-2.3}$ & $-0.28^{+0.59}_{-0.41}$ & $\left(-6.5 ^{+6.1}_{-6.3}\right)\times 10^{-3}$ & $0.92^{+0.05}_{-0.05}$\\
         4 & $12.2^{+0.2}_{-0.2}$ & $12.6^{+0.7}_{-1.1}$ & $0.57^{+0.49}_{-0.38}$ & $1.1^{+0.2}_{-0.4}$ & $4.7^{+3.6}_{-2.3}$ & $-0.51^{+0.56}_{-0.26}$ & $\left(1.4 ^{+7.7}_{-7.2}\right)\times 10^{-3}$ & $1.08^{+0.07}_{-0.07}$\\
         5 & $12.0^{+0.2}_{-0.4}$ & $12.4^{+0.6}_{-1.0}$ & $1.06^{+0.32}_{-0.43}$ & $0.9^{+0.4}_{-0.5}$ & $5.8^{+3.0}_{-2.2}$ & $-0.62^{+0.26}_{-0.14}$ & $\left(2.2 ^{+8.6}_{-8.1}\right)\times 10^{-3}$ & $0.85^{+0.08}_{-0.07}$\\
         \hline
         \multicolumn{9}{c}{\textit{Planck} (CIB corrected)} \\
         \hline
         0 & $12.1^{+0.3}_{-0.3}$ & $12.8^{+0.7}_{-1.3}$ & $0.68^{+0.33}_{-0.36}$ & $1.2^{+0.2}_{-0.4}$ & $3.9^{+4.1}_{-1.9}$ & $-0.46^{+0.67}_{-0.30}$ & $\left(-9.2 ^{+7.4}_{-7.2}\right)\times 10^{-3}$ & $0.98^{+0.06}_{-0.06}$\\
         1 & $12.2^{+0.3}_{-0.3}$ & $13.0^{+0.8}_{-1.5}$ & $0.61^{+0.32}_{-0.40}$ & $1.3^{+0.2}_{-0.1}$ & $1.8^{+3.0}_{-1.0}$ & $-0.59^{+0.52}_{-0.18}$ & $\left(-3.4 ^{+1.2}_{-1.2}\right)\times 10^{-2}$ & $1.31^{+0.10}_{-0.10}$\\
         2 & $11.9^{+0.3}_{-0.5}$ & $13.0^{+0.6}_{-1.2}$ & $0.84^{+0.18}_{-0.26}$ & $1.2^{+0.1}_{-0.1}$ & $0.6^{+0.6}_{-0.3}$ & $-0.65^{+0.28}_{-0.13}$ & $\left(-5.2 ^{+7.0}_{-6.8}\right)\times 10^{-3}$ & $0.88^{+0.06}_{-0.06}$\\
         3 & $12.2^{+0.2}_{-0.3}$ & $12.7^{+0.6}_{-0.9}$ & $0.73^{+0.35}_{-0.49}$ & $1.1^{+0.1}_{-0.2}$ & $2.2^{+3.9}_{-1.0}$ & $-0.50^{+0.56}_{-0.27}$ & $\left(-8.1 ^{+6.5}_{-6.5}\right)\times 10^{-3}$ & $0.92^{+0.05}_{-0.06}$\\
         4 & $12.2^{+0.2}_{-0.4}$ & $12.4^{+0.6}_{-1.1}$ & $0.86^{+0.30}_{-0.50}$ & $1.0^{+0.2}_{-0.4}$ & $5.2^{+3.3}_{-1.9}$ & $-0.86^{+0.21}_{-0.11}$ & $\left(7.4 ^{+7.5}_{-7.3}\right)\times 10^{-3}$ & $1.09^{+0.07}_{-0.07}$\\
         5 & $12.1^{+0.2}_{-0.4}$ & $12.2^{+0.6}_{-0.9}$ & $1.91^{+0.30}_{-0.56}$ & $0.8^{+0.3}_{-0.4}$ & $5.2^{+3.2}_{-1.9}$ & $-0.79^{+0.31}_{-0.14}$ & $\left(2.5 ^{+8.4}_{-8.8}\right)\times 10^{-3}$ & $0.85^{+0.08}_{-0.07}$\\
         \hline
    \end{tabular}
    \caption{Best-fit HOD parameters for the SPT and \textit{Planck} regions.}
    \label{tab:HOD_results}
\end{table*}

\bsp	
\label{lastpage}
\end{document}